\documentclass{emulateapj}

\usepackage{subfigure,epsfig,longtable,wasysym,amssymb,url}

\def\aap{A\&A}

\def\apjs{ApJS}
\def\apjl{ApJL}

\begin{document}

\title{The Feeding Zones of Terrestrial Planets and Insights into Moon Formation}

\author
{Nathan A. Kaib\altaffilmark{1,3} \& Nicolas B. Cowan\altaffilmark{2,3}}

\altaffiltext{1}{Department of Terrestrial Magnetism, Carnegie Institution for Science, 5241 Broad Branch Road, NW, Washington, DC 20015, USA}
\altaffiltext{2}{Department of Physics \& Astronomy, Amherst College, AC\# 2244, Amherst, MA 01002, USA}
\altaffiltext{3}{Center for Interdisciplinary Exploration and Research in Astrophysics (CIERA) and Department of Physics and Astronomy, Northwestern University, 2131 Tech Drive, Evanston, Illinois 60208, USA}

\begin{abstract}

The final stage of terrestrial planet formation consists of several hundred approximately lunar mass bodies accreting into a few terrestrial planets. This final stage is stochastic, making it hard to predict which parts of the original planetesimal disk contributed to each of our terrestrial planets. Here we present an extensive suite of terrestrial planet formation simulations that allows quantitative analysis of this process. Although there is a general correlation between a planet's location and the initial semi-major axes of its constituent planetesimals, we concur with previous studies that Venus, Earth, and Mars analogs have overlapping, stochastic feeding zones. We quantify the feeding zone width, $\Delta a$, as the mass-weighted standard deviation of the initial semi-major axes of the planetary embryos and planetesimals that make up the final planet. The size of a planet's feeding zone in our simulations does not correlate with its final mass or semi-major axis, suggesting there is no systematic trend between a planet's mass and its volatile inventory. Instead, we find that the feeding zone of any planet more massive than $0.1M_\oplus$ is roughly proportional to the radial extent of the initial disk from which it formed: $\Delta a \approx 0.25(a_{\rm max}-a_{\rm min})$, where $a_{\rm min}$ and $a_{\rm max}$ are the inner and outer edge of the initial planetesimal disk. These wide stochastic feeding zones have significant consequences for the origin of the Moon, since the canonical scenario predicts the Moon should be primarily composed of material from Earth's last major impactor (Theia), yet its isotopic composition is indistinguishable from Earth. In particular, we find that the feeding zones of Theia analogs are significantly more stochastic than the planetary analogs. Depending on our assumed initial distribution of oxygen isotopes within the planetesimal disk, we find a $\sim$5\% or less probability that the Earth and Theia will form with an isotopic difference equal to or smaller than the Earth and Moon's. In fact we predict that every planetary mass body should be expected to have a unique isotopic signature. In addition, we find paucities of massive Theia analogs and high velocity moon-forming collisions, two recently proposed explanations for the Moon's isotopic composition. Our work suggests that there is still no scenario for the Moon's origin that explains its isotopic composition with a high probability event.

\end{abstract}

\section{Introduction}
An outstanding question in planetary science is the degree to which a planet's volatile inventory and isotopic composition is related to its size and location.  In other words, how deterministic is terrestrial planet formation? In this paper we use ensembles of planet formation simulations to study the statistics of terrestrial planet feeding zones. Insofar as the initial planetesimal disks have radially non-uniform volatile content and isotopic composition, the width of feeding zones may be used as a proxy for the final water inventory and isotopes of planets.   

The water inventory of rocky worlds is important: liquid water is the definition of habitability \citep{Kasting_1993}, but planets with too much surface water will have no exposed continents, and hence will not benefit from the silicate weathering thermostat \citep{Abbot_2012}.  The combined effects of erosion, isostacy, and a substantial mantle water reservoir mean that planets with water mass fractions $<10^{-3}$ will not inundate their surface \citep{Cowan_2014}. Insofar as a terrestrial planet's water is delivered in the normal course of planet formation \citep{ray04}, the statistics of planetary feeding zones affect the statistics of habitable planets. 

\subsection{Theia \& the Moon-forming Impact}

Furthermore, understanding the variance in isotopic composition predicted from terrestrial planet formation is key to deciphering the origin of our Moon. In the canonical giant impact hypothesis for the Moon's origin, a Mars-mass body (named Theia) strikes the proto-Earth in a glancing impact at the tail end of the giant impact phase of terrestrial planet formation \citep{hartdav75, camward76}. This impact throws material into orbit around Earth, which eventually accretes into the moon we have today. Using high-resolution smooth-particle hydrodynamic simulations, such a collision has been demonstrated to yield a Lunar-mass satellite that is depleted in iron and also gives the Earth-Moon system approximately the angular momentum observed today \citep{canup01}. In such collisions, however, the moon-forming accretion disk around the Earth is largely composed of material from the impactor rather than the proto-Earth \citep{canup04}. This presents a potential problem for the giant impact hypothesis because the oxygen isotope composition of lunar samples has been found to be virtually identical to terrestrial rocks ($\Delta^{17}$O$<0.016$\permil) \citep{wiechert01}. In contrast, meteorites from Mars and Vesta are distinctly different from the Earth, with $\Delta^{17}$O values of 0.32\permil~ and -0.28\permil, respectively \citep{franch99, clay96}. More recent work hints at a slight difference between the Moon and the Earth of $\Delta^{17}$O$\simeq0.012$\permil, but the fact remains that the Moon is much more isotopically similar to Earth than samples from any other large object in the solar system \citep{her14}.

Because the formation of the Moon requires a relatively low-velocity impact between Theia and the proto-Earth, it had previously been argued that Theia likely resided in an orbit very similar to the proto-Earth and may have formed from a similar region of the planetesimal disk \citep{wiechert01}. If Theia and Earth formed from the same pool of planetesimals, they may have a similar composition, while Mars and Vesta have distinctly different oxygen signatures, owing to their different feeding zones in the planetesimal disk. However, the late stages of terrestrial planet formation are not a local process. Numerical studies of the final giant impact phase of terrestrial planet formation have conclusively demonstrated that the feeding zones of terrestrial planets are quite large and that most bodies undergo substantial and stochastic radial migration  \citep{cham01, ray04, ray05, obrien06, ray06, ray09, fischcies14}. As a result, there is no reason to expect that Earth and its final impactor formed from the same zones of the protoplanetary disk. Hence, the identical composition of the Moon seems to rely on a potentially improbable coincidence. 

In light of this, there have been recent searches for alternative scenarios to explain the similarities between lunar and terrestrial rocks. Instead of the canonical Mars-mass impactor, \citet{canup12} explores the outcomes of collisions between a $\sim$0.5 M$_{\oplus}$ Theia and a $\sim$0.5 M$_{\oplus}$ proto-Earth. For certain impact velocities, the impactor makes nearly equal contributions to the Earth and the moon-forming disk, resulting in an isotopically similar Earth and Moon. This scenario tends to give the Earth-Moon too much angular momentum, however. To alleviate this problem, \citet{cukstew12} suggest that the system's angular momentum could have decreased by up to a factor of 2 when it passed through a solar evection resonance; this allows for a higher impact velocity between the proto-Earth and Theia as well. If the proto-Earth had a rotation period of 2--3 hours, \citet{cukstew12} show that high velocity impacts can yield a lunar composition dominated by the proto-Earth. Similarly, \citet{reufer12} finds that a hit-and-run collision at a slightly higher impact velocity can diminish Theia's contribution to the Moon's composition.

Another hypothesis for the isotopic similarities of the Earth and Moon is that they mixed immediately after the impact \citep{pahlstev07}. The energy from the moon-forming impact would have vaporized portions of the outer Earth and generated a vapor accretion disk initially comprised of the impactor. The disk and outer Earth could have mixed and isotopically equilibrated in the 10$^{2-3}$ years that the disk was in vapor form. During the time required for mixing, however, the outer portion of the disk would have already cooled and begun accreting into the Moon. In addition, in this scenario, one would expect refractory elements to exhibit stronger isotopic differences between the Earth and Moon since they would have condensed into solids faster, yet such a trend is not seen \citep{zhang12, arm12}. 

We may not need to abandon the canonical giant impact hypothesis, however. It is generally presumed that isotopically similar Theia and Earth are improbable outcomes of terrestrial planet formation, but the probability has not yet been quantified. While terrestrial planet formation simulations have recently been used to constrain the original orbit, timing, and collision statistics for Theia \citep{quarles14, jac14, jacmorb14}, few large statistical studies of the potential compositions and feeding zones of terrestrial planets (and their impactors) have been undertaken \citep{izi13, fischcies14}. Because the feeding zones of terrestrial planets can be so large, it is intuitively plausible that Earth acquired the disk-averaged value of $\Delta^{17}$O due to the planet's high mass \citep{ozima07}. If this were the case, then Mars-mass bodies like Theia would likely possess similar $\Delta^{17}$O values, with Mars and Vesta simply being counter-examples; the unaltered canonical giant impact hypothesis could still be a viable explanation for the Moon's origin.

With this in mind, we have revisited the problem of late stage terrestrial planet formation. We have performed 150 different simulations modeling the formation of our terrestrial planets under different initial conditions. We use our simulations to make a statistical study of the relationship between a planet's final properties---mass and semimajor axis---and its accretion history and feeding zone. Such relationships can then help us discern how often, if ever, the accretion in our simulations results in isotopically similar Earth and Theia analogs, but still allows for an isotopically distinct Mars. This will enable us to estimate the probability that Theia had an Earth-like composition and will help assess the plausibility of the canonical giant impact hypothesis. 

The large number of simulations we have performed also gives us new insights into the general process of the giant impact phase of terrestrial planet formation. In particular, we can statistically study the relationship between the final location of a planet and the location of its accretionary feeding zone within the original planetesimal disk. With 150 simulations, we will also be able to better quantify the level of stochasticity of these feeding zones and how this may affect the delivery of distant water-rich planetesimals to the terrestrial planets.

Our work is organized into the following sections: Section \ref{sec:nummeth} describes the numerical methods and initial conditions employed in our simulations. Following this, we have five main sections of results. The first (Section \ref{sec:trends}) presents a general overview of the relationship between the final mass and orbit of a terrestrial planet and its feeding zone in the initial planetesimal disk; Section~\ref{sec:genrelations} specifically discusses the feeding zones for analogs of Solar System planets. Then in \ref{sec:theiacomp}, we employ several different initial hypothetical distributions of $\Delta^{17}$O in our simulations, and we determine how often Earth-like Thieas are formed alongside isotopically distinct Mars analogs. With the work of \citet{canup12} and \citet{cukstew12} in mind, we next use our simulations to quantify the mass and impact velocity distributions of Theia analogs in Section \ref{sec:theiamass}. In Section \ref{sec:venuscomp} we then make predictions for the isotopic composition of Venus relative to Earth based on the $\Delta^{17}$O distributions employed in the study of Theia's composition. We summarize the conclusions of our work in Section \ref{sec:summ}.

\section{Numerical Methods}\label{sec:nummeth}

We have performed three ensembles of 50 simulations modeling the giant impact phase of terrestrial planet formation in the solar system with the MERCURY hybrid integration package \citep{cham99}. Each ensemble makes different assumptions about the distribution of planetesimals as well as the orbits of Jupiter and Saturn. The first ensemble, ``circular Jupiter and Saturn'' (CJS), assumes that Jupiter and Saturn begin with nearly circular orbits ($e<0.01$) on their present-day semimajor axes. We begin with 100 self-interacting embryos orbiting between 0.5 and 4.0 AU on nearly circular ($e<0.01$), nearly coplanar ($i<1^{\circ}$) orbits. These embryos' semimajor axes are spaced evenly, and their masses decrease with semimajor axis to yield a $a^{-3/2}$ surface density profile. In addition, each simulation includes 1000 smaller planetesimals that do not gravitationally interact with each other but do interact with all of the other simulation bodies. These planetesimals are given a fixed mass, and their initial semimajor axis distribution is adjusted to yield a $a^{-3/2}$ surface density profile as well. Like the embryos, the planetesimals are initially placed on nearly circular, coplanar orbits. Arguments of perihelion, longitudes of ascending node, and mean anomalies are randomly generated from uniform distributions for all bodies. Our disk mass of $\sim$5 M$_{\oplus}$ is split evenly between embryos and planetesimals. Finally, we set the radii of embryos and planetesimals by assuming a bulk density of 3 g/cm$^3$.

\citet{ray09} demonstrated that terrestrial planet formation simulations with an initially circular Jupiter and Saturn consistently produce planets near 1.5 AU that are much more massive than the real Mars (Figure~\ref{analogs_cjs}). Initializing Jupiter and Saturn on eccentric orbits quickly truncates the planetesimal disk via secular resonances, resulting in Martian analogs more akin to the real Mars (Figure~\ref{analogs_ejs}). We therefore include a second set of simulations, ``eccentric Jupiter and Saturn'' (EJS). These simulations start Jupiter and Saturn with moderately eccentric orbits ($e=0.1$) at their present-day semimajor axes. The distributions of embryos and planetesimals in EJS is identical to that employed in CJS.

Another failure of many terrestrial planet formation simulations is that they fail to form the two most massive planets between 0.7 and 1.0 AU as our own solar system has. However, the radial concentration of planetary mass seen in our own solar system can be replicated if terrestrial planet formation begins from a narrow annulus (Figure~\ref{analogs_ann}) \citep{mor08, hans09}. We mimic this in our last ensemble, ``annulus'' (ANN). As in CJS, we assume nearly circular coplanar orbits for Jupiter and Saturn ($e<0.01$, $i<1^{\circ}$). However, in this ensemble all of our embryos are confined to initial semimajor axes in an annulus between 0.7 and 1.0 AU. This also crudely replicates the planet formation scenario put forth in the Grand Tack Model \citep{walsh11}. The ANN simulations contain 400 embryos, each with a mass of 0.005 M$_{\oplus}$, on nearly circular, coplanar orbits with random mean anomalies, longitudes of ascending node, and arguments of perihelion; no planetesimals are present. This final ensemble has identical starting conditions to those of \citet{hans09}. 

We choose these three sets of initial conditions because they are similar to those employed in numerous popular past studies of terrestrial planet formation, and we wish to see how objects' accretion histories vary under different assumptions. Each set has shortcomings and there are many other potential initial conditions that we do not explore [e.g. ][]\citep{ray09, izi14}.  It is beyond the scope of the present work to advocate one set of initial conditions over another. Rather, we employ these different initial conditions to explore the possible parameter space of the solar system's terrestrial planet formation.

For all simulations, we integrate orbits with a time step of 6 days for 200 Myrs. Particles are removed from simulations if they pass within 0.1 AU of the Sun or if they go beyond 100 AU. The computing time to complete this large suite of simulations is quite significant, and we will make our simulations publicly available at \url{http://nathankaib.com}.

\section{Results}
\subsection{Feeding Zone Trends}\label{sec:trends}

We first consider whether there are general trends between the location and width of a planet's feeding zone and its final properties, namely semi-major axis and mass.  While previous works have studied feeding zone stochasticity and correlation with final orbit \citep[e.g., ][]{cham01, ray04, fischcies14}, our large number of simulations and large particle number per simulation allow us to search for general statistical trends that may have been less obvious in past work.

\begin{figure}
\centering
\includegraphics[scale=.46]{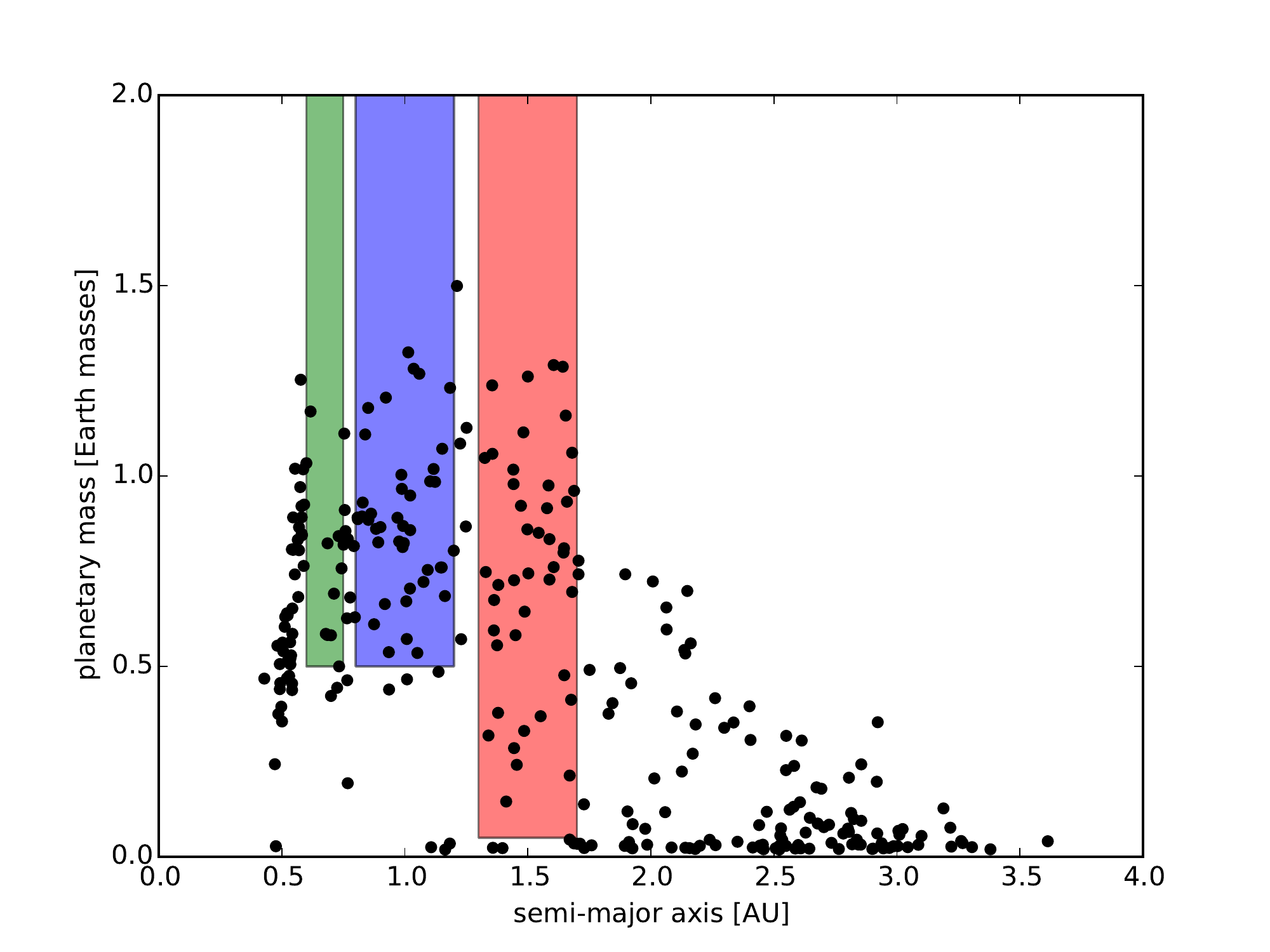}
\caption{The final mass and semi-major axis for the CJS ensemble (initially circular orbits for Jupiter and Saturn).  The colored boxes indicate analogs of Venus (green), Earth (blue), and Mars (red).}\label{analogs_cjs}
\end{figure}

\begin{figure}
\centering
\includegraphics[scale=.46]{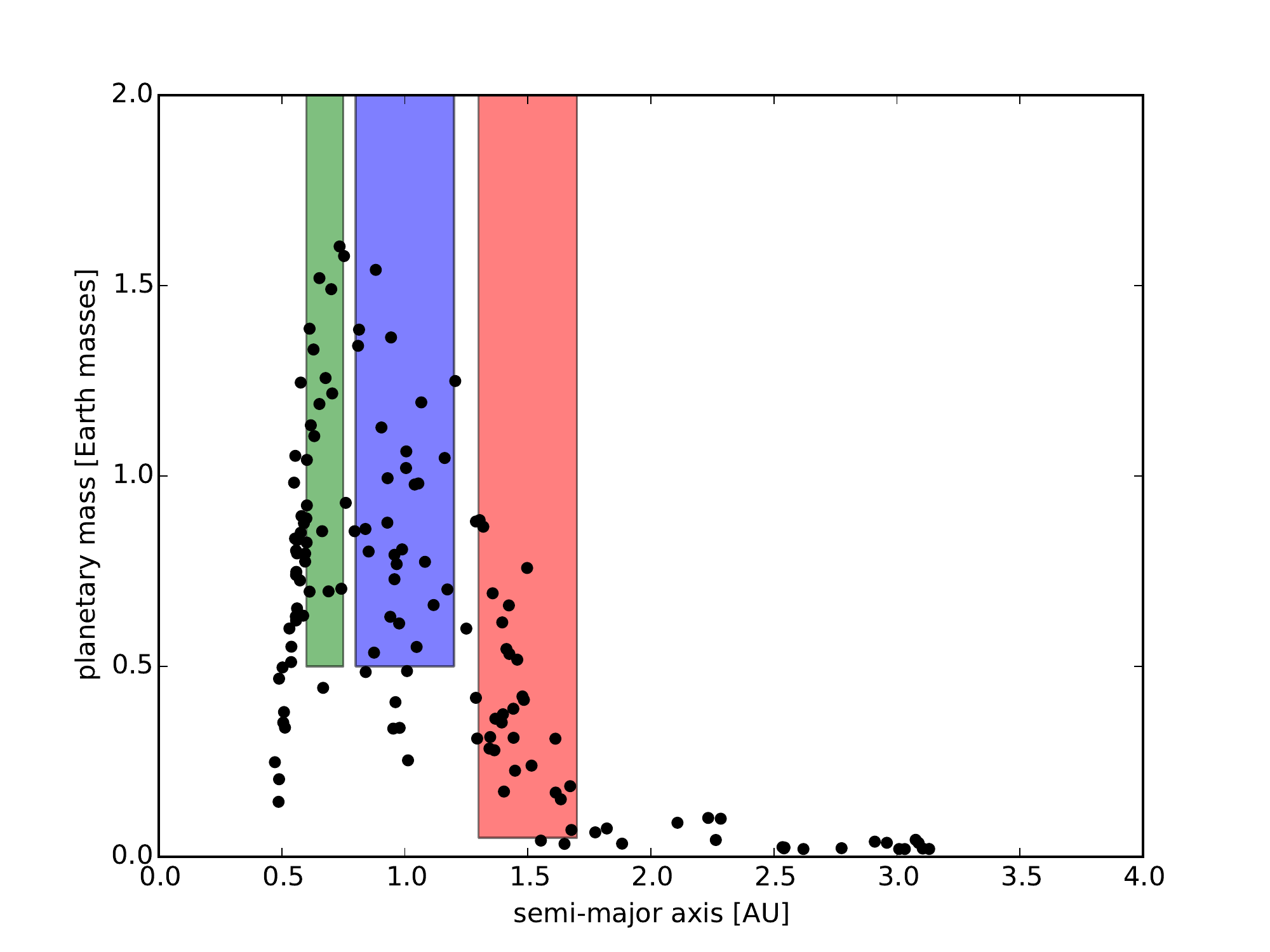}
\caption{The final mass and semi-major axis for the EJS ensemble (initially eccentric orbits for Jupiter and Saturn).  The colored boxes indicate analogs of Venus (green), Earth (blue), and Mars (red).}\label{analogs_ejs}
\end{figure}

\begin{figure}
\centering
\includegraphics[scale=.46]{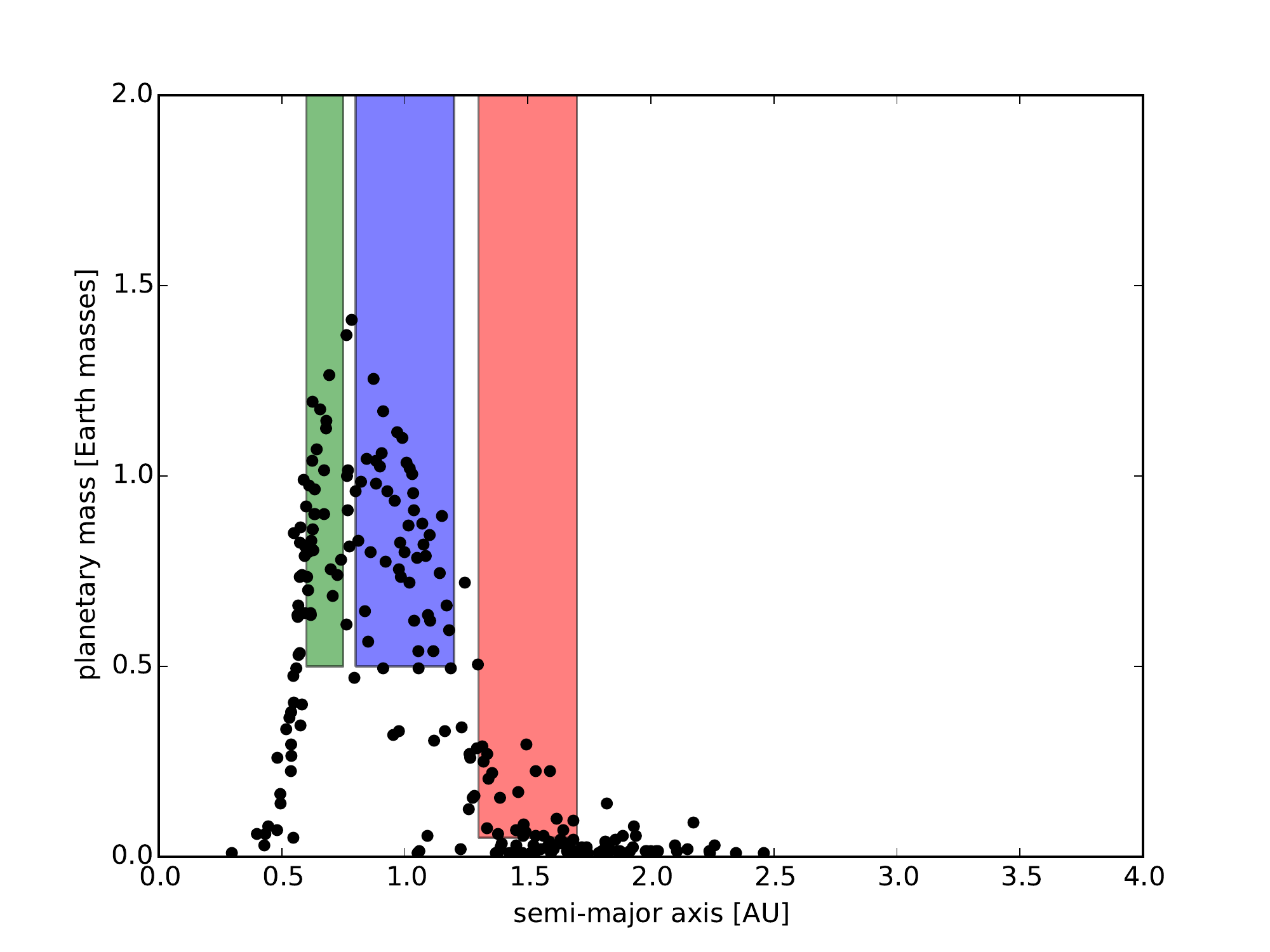}
\caption{The final mass and semi-major axis for the ANN ensemble (0.7--1.0~AU annulus of planetary embryos).  The colored boxes indicate analogs of Venus (green), Earth (blue), and Mars (red).}\label{analogs_ann}
\end{figure}

We pay special attention to planets with attributes similar to Solar System terrestrial planets and have defined categories for Venus, Earth and Mars analogs in our simulations. Venus analogs have final semimajor axes between 0.6 and 0.75 AU and masses greater than 0.5 M$_{\oplus}$. Similarly, planets are designated as Earth analogs if 0.8 $< a <$ 1.2 AU and $m>0.5$ M$_{\oplus}$, while Mars analogs are defined as $1.3 < a < 1.7$ AU and $m>0.05$ M$_{\oplus}$. The ranges of $a$ and $m$ employed to define our analogs are quite liberal, but this must be done to generate a statistically useful sample of analogs of each class (Figures~\ref{analogs_cjs}--\ref{analogs_ann}). 

\begin{figure}
\centering
\includegraphics[scale=.46]{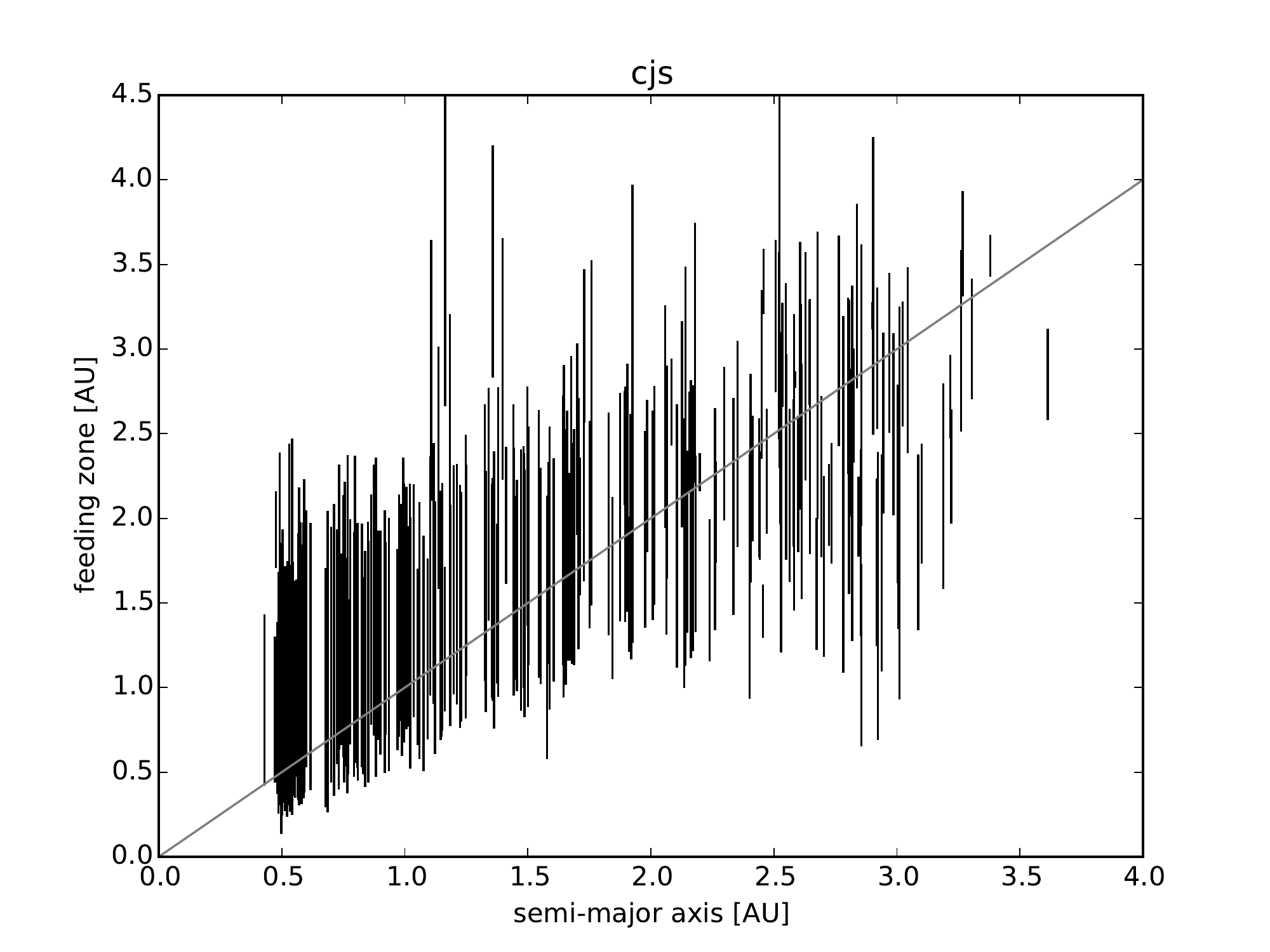}
\caption{The location (mass-weighted mean) and width (mass-weighted standard deviation) of a planet's feeding zone is plotted vs.\ its final semi-major axis. The diagonal gray line denotes a 1:1 correspondence between planetesimals and the planets final location.}\label{mwm_feeding_zone_vs_semi_major_CJS}
\end{figure}

\begin{figure}
\centering
\includegraphics[scale=.46]{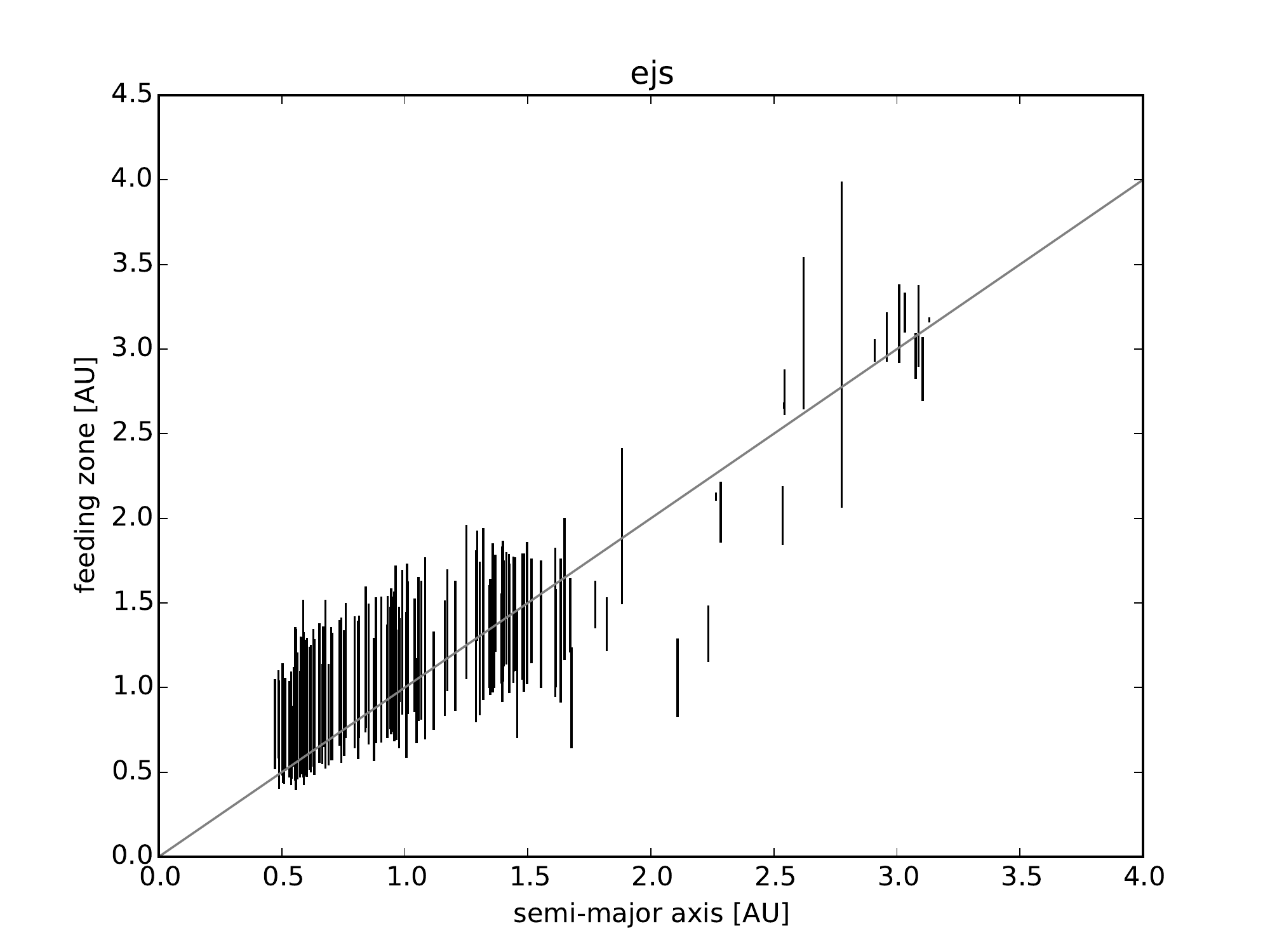}
\caption{The location (mass-weighted mean) and width (mass-weighted standard deviation) of a planet's feeding zone is plotted vs.\ its final semi-major axis. The diagonal gray line denotes a 1:1 correspondence between planetesimals and the planets final location.}\label{mwm_feeding_zone_vs_semi_major_EJS}
\end{figure}

\begin{figure}
\centering
\includegraphics[scale=.46]{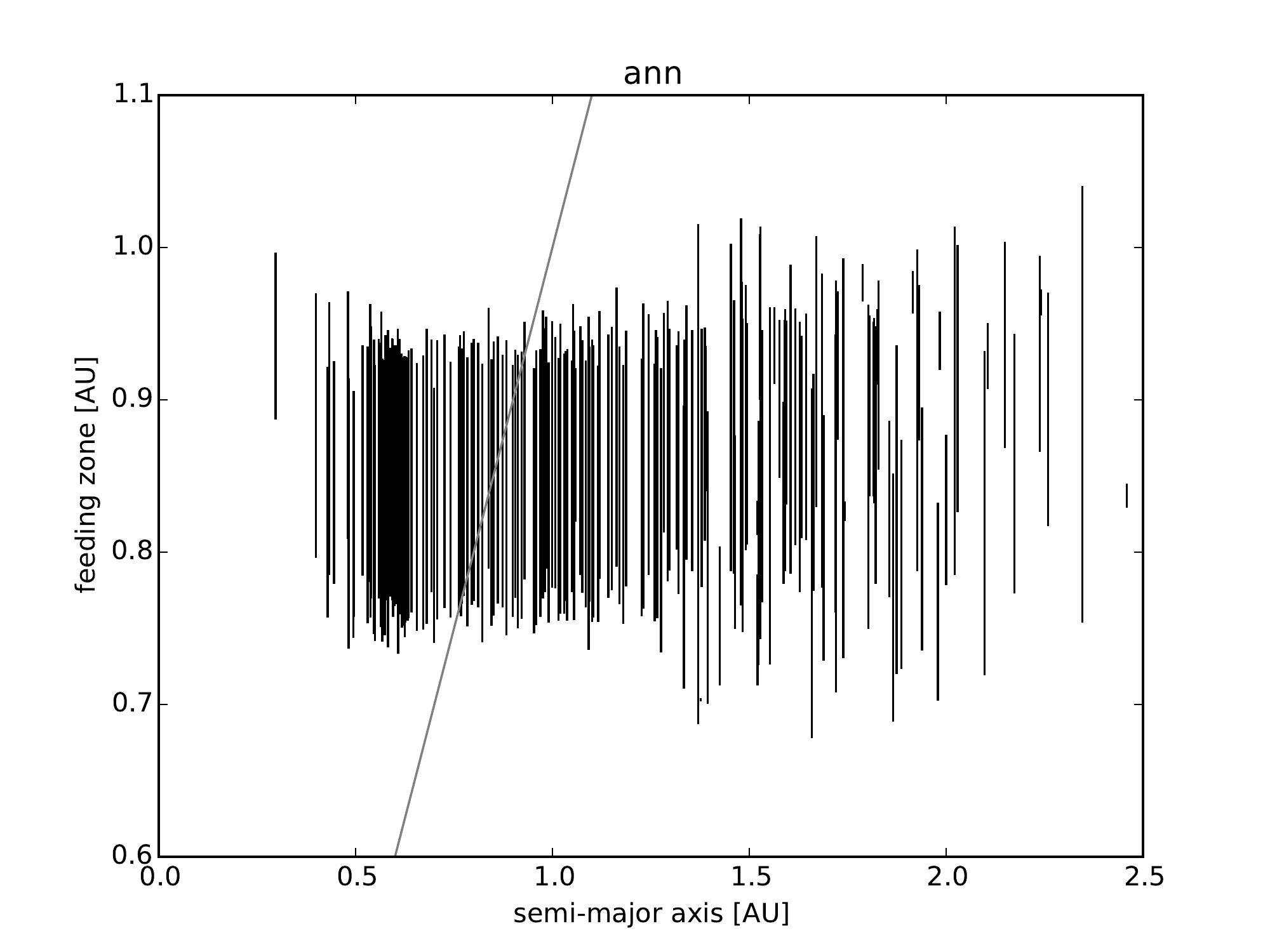}
\caption{The location (mass-weighted mean) and width (mass-weighted standard deviation) of a planet's feeding zone is plotted vs.\ its final semi-major axis. The diagonal gray line denotes a 1:1 correspondence between planetesimals and the planets final location.}\label{mwm_feeding_zone_vs_semi_major_ANN}
\end{figure}

In Figures~\ref{mwm_feeding_zone_vs_semi_major_CJS}, \ref{mwm_feeding_zone_vs_semi_major_EJS}, and \ref{mwm_feeding_zone_vs_semi_major_ANN} we plot the planetary feeding zones vs.\ final semi-major axes for our three ensembles of simulations. The location of a planet's feeding zone is the mass-weighted mean semi-major axis of its constituent planetesimals; the feeding zone width is the mass-weighted standard deviation of its constituent planetesimals' semi-major axes. 

\begin{figure}
\centering
\includegraphics[scale=.46]{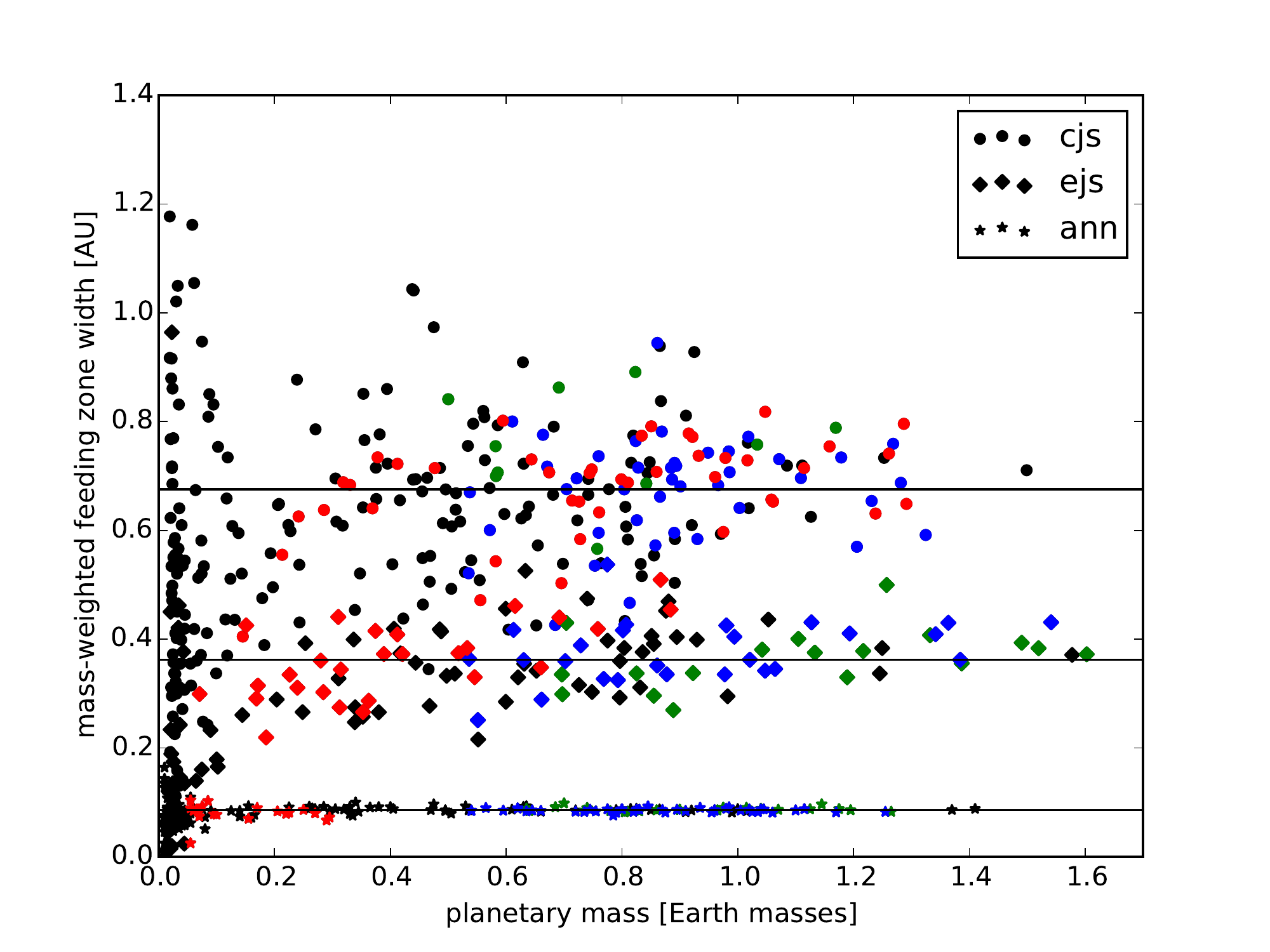}
\caption{The mass-weighted standard deviation of initial semi-major axes is plotted vs.\ its mass. The symbol shape denotes the ensemble of simulations from which the planets were drawn. For each ensemble, the median feeding zone width for planets larger than $0.1M_\oplus$ is shown as a black line. The colored symbols are analogs of Venus (green), Earth (blue), and Mars (red).}\label{mwv_feeding_zone_vs_total_all}
\end{figure}

\begin{figure}
\centering
\includegraphics[scale=.46]{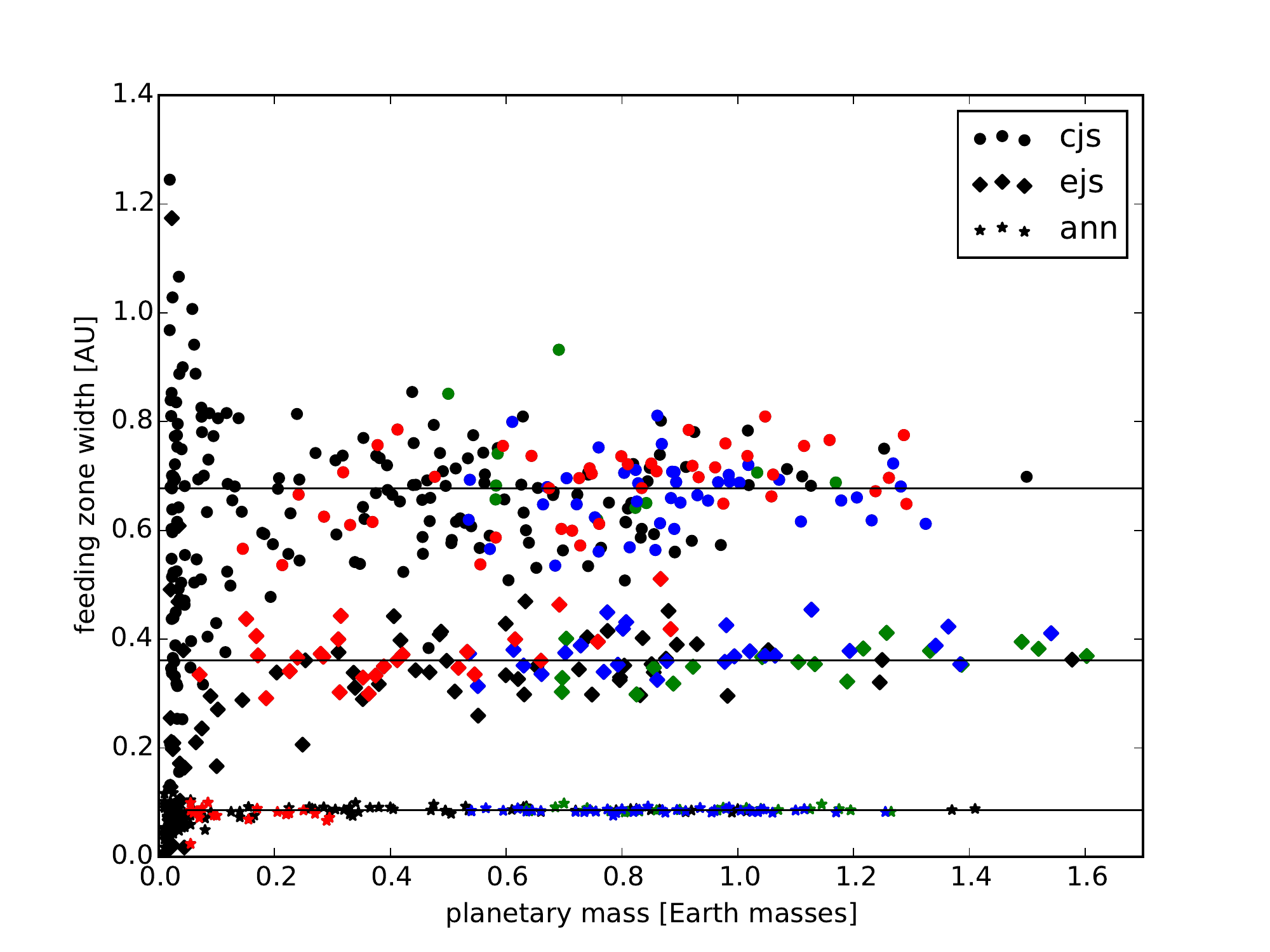}
\caption{The (unweighted) standard deviation of initial semi-major axes is plotted vs.\ its mass. The symbol shape denotes the ensemble of simulations from which the planets were drawn. For each ensemble, the median feeding zone width for planets larger than $0.1M_\oplus$ is shown as a black line. The colored symbols are analogs of Venus (green), Earth (blue), and Mars (red).}\label{std_feeding_zone_vs_total_all}
\end{figure}

Planets that end up close to their star tend to form from material close to the star, and vice versa, but the relation is not 1:1 because the initial disk has a finite extent and the material spreads out.  Moreover, individual planets have feeding zones that span a large fraction of the initial disk, and the width of the feeding zone does not exhibit a strong trend with final-semi-major axis.   

Nor does the width of a planet's feeding zone correlate with its mass (Figure~\ref{mwv_feeding_zone_vs_total_all}).  In fact, while low-mass planets have a variety of feeding zone widths, planets more massive than $0.1M_\oplus$ have approximately the same size feeding zone for each ensemble. This behavior is even more striking if one considers the unweighted standard deviation of initial semi-major axes, as shown in Figure~\ref{std_feeding_zone_vs_total_all}.  In other words, planets accrete planetesimals from a fairly consistent feeding zone, but the few accreted planetary embryos tend to be more of a crapshoot; this has important consequences when we consider Theia analogs, below. 

The constant feeding zone size of terrestrial planets differs between our three ensembles: $\Delta a_{\rm CJS} = 0.7(1)$~AU, $\Delta a_{\rm EJS} = 0.36(7)$~AU, and $\Delta a_{\rm ANN} = 0.086(5)$~AU (the values for the un-weighted standard deviation are $\Delta a_{\rm CJS} = 0.68(8)$~AU, $\Delta a_{\rm EJS} = 0.36(5)$~AU, and $\Delta a_{\rm ANN} = 0.086(5)$~AU).  The radial mixing statistic \citep{cham01,quin14}, $S_R$, is 0.57, 0.39, and 0.27 for the CJS, EJS, and ANN ensembles, respectively.  

The feeding zone widths are proportional to the initial disk size from which the planets form. The CJS ensemble starts with planetesimals and embryos at 0.5--4.0~AU for a width of 3.5~AU.  The EJS ensemble effectively has an initial annulus of 0.5--2.5~AU (width of 2.0~AU) since planetesimals beyond 2.5~AU are rapidly ejected by resonances with the eccentric gas giants. Finally, the 0.7--1.0~AU annulus configuration has a width of 0.3~AU.  Given these three ensembles, it therefore appears that the width of a planet's feeding zone is 20--30\% the width of the initial planetesimal disk.

\subsection{Feeding Zones of Solar System Analogs}\label{sec:genrelations}

\begin{figure}
\centering
\includegraphics[scale=.43]{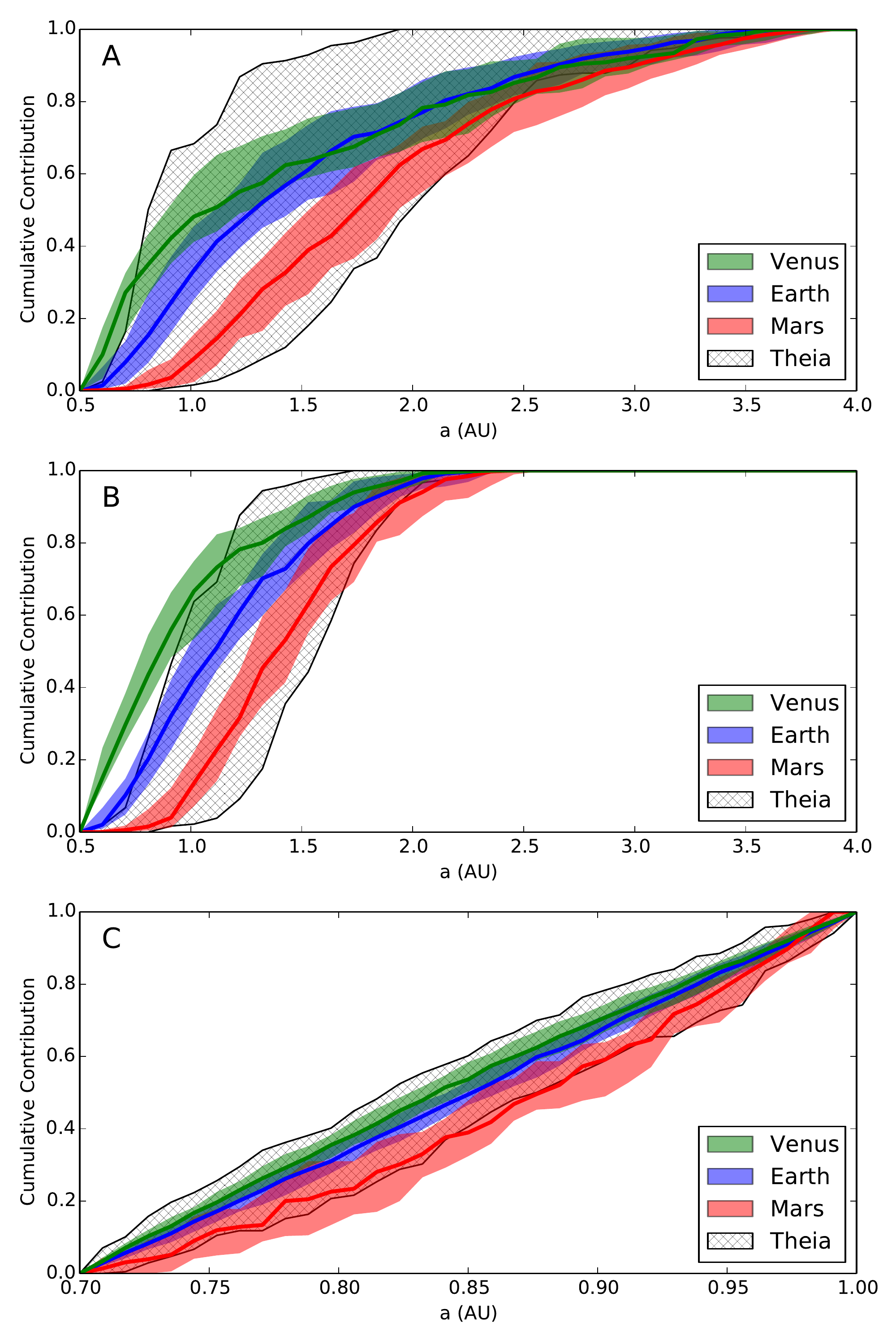}
\caption{The relative contribution of each region of the planetesimal disk to the final compositions of Venus, Earth, Mars, and Theia analogs are show with green, blue, red, and hashed CDFs, respectively. The median CDF of each class of analogs is shown with a solid line, and the 1-$\sigma$ range of CDFs seen amongst our simulations for each analog class is marked with the shaded region. Panels {\bf A}, {\bf B}, and {\bf C} show the results from the CJS, EJS, and ANN simulations, respectively.}\label{fig:cdfs}
\end{figure}

For each analog planet we have gone back and looked at the relative contribution that planetesimals and embryos in our initial disk make toward its construction. Weighing all bodies by their mass, we are then able to see which regions of the protoplanetary disk are sampled most heavily during the construction of Venus, Earth, and Mars analogs. The results of this analysis are shown in Figure \ref{fig:cdfs}; panel A shows the CJS simulations. We see that Mars analogs in these simulations have a distinct composition relative to Venus and Earth analogs. In the previous section we saw that planets forming in a disk tend to all have the same size feeding zone, so it stands to reason that Earth and Venus (0.3~AU apart) will have more overlap than Earth and Mars (0.5~AU apart). Although there is a stochastic nature to terrestrial planet formation, this plot shows that Mars analogs consistently form from material orbiting further from the Sun. This is less true for Venus and Earth analogs: although on average Earth analogs form from more distant material than Venus analogs, the two planets' cumulative distribution functions overlap significantly. In addition, we find that analogs of all three terrestrial planets typically receive comparable contributions of material beyond 2.5 AU, which is presumably water-rich \citep{hay81, ray04}. 

We also search our simulations for Theia analogs, which we define to be the last body with $m>0.1$ M$_{\oplus}$ that struck an Earth analog. (Previous studies have found that our moon could be formed with impactors less massive than 0.1 M$_{\oplus}$, but it is clear from Figures~\ref{mwv_feeding_zone_vs_total_all} and \ref{std_feeding_zone_vs_total_all} that our simulation's mass resolution limit strongly affects these smaller bodies' feeding zones.) We see in Figure \ref{fig:cdfs}A that the composition of Theia analogs varies much more than that of planetary analogs.  In some cases Theia analogs are built from material more distant than Mars analogs, while in other cases they have compositions dominated by material nearer to the Sun than Venus analogs' compositions. Part of the reason for this variation is that Theia analogs are simply smaller bodies on average. Thus, their accretion history samples fewer regions of the disk, and an ensemble of Theia CDF curves displays more variance than planetary CDF curves. However, by definition Theia is a planetary mass body that is quasi-stable at best. Eventually it becomes unstable and strikes the Earth. Theia's very wide CDF swath in Figure \ref{fig:cdfs}A suggests that before impact, Theia can be quasi-stable interior to the Earth (implying a composition dominated by inner disk material) as often as it is quasi-stable exterior to the Earth (which would imply a greater fraction of outer disk material in its composition). This fact also helps explain the large compositional variance of Theia analogs.

The EJS simulations are shown in Figure \ref{fig:cdfs}B. In these simulations, material from outside 2.5 AU makes nearly no contribution to the compositions of the planets. This is because the eccentricities of these distant planetesimals are rapidly excited via strong secular and resonant perturbations from Jupiter, which quickly leads to their ejection before they can be accreted by the growing terrestrial planets. In addition, we see that Venus, Earth, and Mars analogs all have narrower CDF swaths with less overlap, indicating a more deterministic outcome of planet formation. However, Theia analogs still display a very large compositional variance compared to the planets, as some Theia analogs are largely formed from material beyond Mars whereas others are mostly formed from material between Venus and Earth.

Finally, we investigate the ANN simulations in Figure \ref{fig:cdfs}C. Here we find that Venus and Earth analogs typically have very similar compositions with material at all distances making nearly identical contributions to both planets. Furthermore, there are less dramatic compositional differences between Earth and Mars analogs in this ensemble compared to CJS and EJS. Mars has only a slight bias toward more distant material, and there is overlap with the CDF swath for Earth analogs. In fact, in 12.5\% of the runs that contain Earth and Mars analogs, the mean location of the Martian feeding zone is actually closer to the Sun than the Earth analog's feeding zone's mean location (in stark contrast to our CJS or EJS simulations). Once again, we see a substantially higher level of variance for the composition of Theia analogs compared to the planets.

\subsection{Oxygen Composition of Earth, Theia, and Mars}\label{sec:theiacomp}

Because of the chaotic nature of the final stage of planet formation, an Earth-like oxygen isotope composition for Theia is considered improbable. While two past studies of the $\Delta^{17}$O values within a single planet formation simulation suggest this is the case, the probability that Theia possessed an Earth-like oxygen isotope composition has never been estimated, even though terrestrial planet formation simulations have the potential to do so \citep{zieg06, pahlstev07}. Given an initial distribution of $\Delta^{17}$O among embryos and planetesimals, it is easy to predict the $\Delta^{17}$O values of the final planets by weighting each accreted body's $\Delta^{17}$O contribution by its mass. This approach is challenging because we have virtually no constraints on the functional form of the initial $\Delta^{17}$O distribution in the solar nebula. Nevertheless, we can test different functional forms and use the difference between Earth's and Mars' fractionation as constraints on our chosen hypothetical distributions. (This approach of employing hypothetical distributions is similar to that taken by \citet{quin14} when studying water delivery during terrestrial planet formation.) With the constraints provided by Earth's and Mars' isotopes, we can then attempt to estimate the probability that the final major impactor with the Earth (Theia) acquires an Earth-like distribution of oxygen isotopes. 

Below we investigate several different simple families of initial distributions of $\Delta^{17}$O on our simulations that contain Earth, Mars, and Theia analogs. To increase the fraction of analyzed simulations, we further relax our criteria for Mars analogs to simply be the first planet orbiting beyond an Earth analog (as defined above). This yields 68 different simulations for analysis (37 from CJS, 16 from EJS and 15 from ANN).

\subsubsection{Linear Distribution}

\begin{figure}
\centering
\includegraphics[scale=.43]{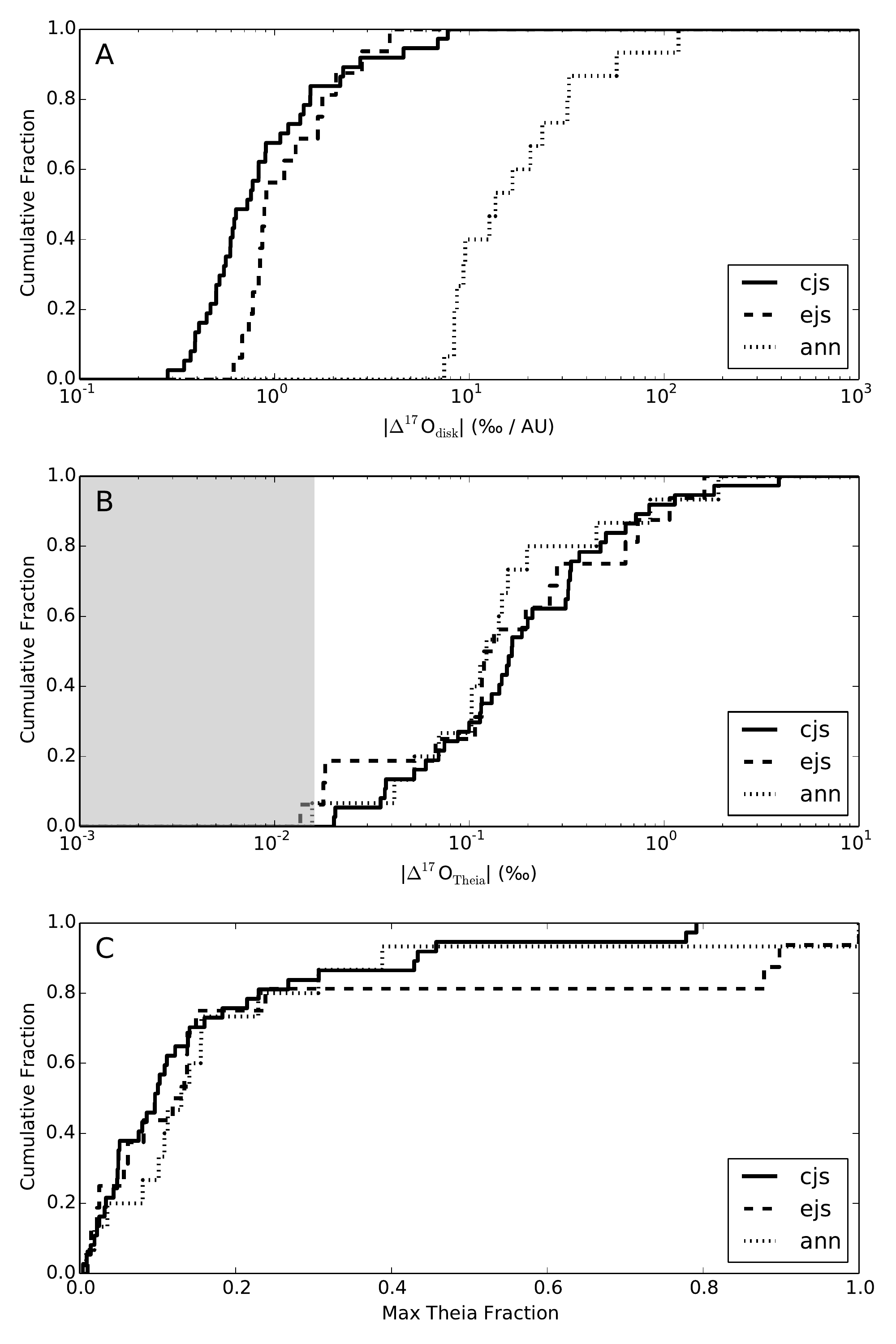}
\caption{{\bf A:} Assuming an initial distribution of $|\Delta^{17}$O$|$ among planetesimals and embryos that varies linearly with heliocentric distance, a histogram of the $|\Delta^{17}$O$|$ distribution slopes is shown for all of our simulations that form Earth, Mars, and Theia analogs. Each distribution is tuned so each simulation yields $|\Delta^{17}$O$|=0.32$\permil~ when comparing the final oxygen isotope compositions of Earth and Mars analogs. {\bf B:} Using the distributions specified in panel A, the $|\Delta^{17}$O$|$ values of Theia analogs in each simulation are calculated and shown here. The shaded regions marks values of $|\Delta^{17}$O$|$ below 0.016\permil. {\bf C:} Given the Theia $|\Delta^{17}$O$|$ values in panel B, the maximum allowable contribution to our Moon's mass is calculated for each Theia analog. The CDFs of these allowable contributions are shown here. CJS, EJS, and ANN simulation results are displayed with the solid, dashed, and dotted CDFs, respectively, in all three panels.}\label{fig:theialine}
\end{figure}

The simplest non-uniform initial distribution we can assign for $\Delta^{17}$O is a linear function in $a$. In each simulation that forms Mars, Earth, and Theia analogs, we impose on our initial planetesimals and embryos a distribution in $\Delta^{17}$O that varies linearly with $a$. In each simulation, the slope and intercept of the distribution is chosen so that the final $\Delta^{17}$O differs by 0.32$\permil$ between Mars and Earth. Thus, each simulation has a unique initial $\Delta^{17}$O distribution tuned to replicate the observed difference between Earth and Mars. The slopes of these distributions are shown in Figure \ref{fig:theialine}A. For our simulations with a disk of planetesimals between 0.5 and 4 AU (CJS and EJS), the $\Delta^{17}$O gradient is of order 1\permil ~per AU. It is much steeper for ANN, since the annulus has a smaller spread in initial semimajor axes and forces Earth and Mars analogs to have feeding zones that overlap more.

With these $\Delta^{17}$O gradients chosen, we can now look at the distribution of $\Delta^{17}$O values for the Theia analogs in each simulation. These are shown in Figure \ref{fig:theialine}B. (We plot the absolute magnitude of $\Delta^{17}$O, since we are more concerned with the difference in $\Delta^{17}$O than its sign.) As can be seen, the difference in $|\Delta^{17}$O$|$ for Earth and Theia is almost always greater than that observed in the real solar system (assuming the Moon is composed of material from Theia \citep{canup04}). The median $|\Delta^{17}$O$|$ value of Theia analogs in the CJS, EJS, and ANN simulations is 0.17, 0.13, and 0.12\permil, respectively. Out of the 37 systems we study in the CJS set, none of them manage to generate a Theia analog with $|\Delta^{17}$O$|$ below 0.016\permil. In the EJS simulations, only 1 of 16 systems manage to produce an Earth-like Theia. Similarly, in the ANN simulations only 1 of 15 systems generate a Theia analog with $|\Delta^{17}$O$|<0.016$\permil. The failure to generate Earth-like Theia analogs in these simulations is surprising. In all but two simulations we fail at generating a Theia analog with an Earth-like oxygen composition in spite of a smooth linear $\Delta^{17}$O designed to produce the observed difference between Earth and Mars. This suggests that either the canonical moon formation scenario is unlikely, or that the initial $\Delta^{17}$O distribution in the solar system was very different from a linear distribution.

The canonical collision scenario for the Moon's origin predicts that at least 70\% of the Moon's material will be derived from Theia \citep{canup04}. More recent alternative collision scenarios successfully decrease that fraction to values of $\sim$5--40\% \citep{cukstew12,reufer12}. From our results in Figure \ref{fig:theialine}B, for each Theia analog we can easily calculate its maximum contribution to the Moon's composition allowed by the measured lunar isotopic composition. The distribution of these allowable contributions is shown in Figure \ref{fig:theialine}C. Here we see that the large majority of our Theia analogs are prohibited from contributing more than $\sim$40\% of the Moon's mass. The median value of the allowable Theia contribution is 9.7\%, 13\%, and 13\% for CJS, EJS, and ANN simulations, respectively.

\begin{figure}
\centering
\includegraphics[scale=.43]{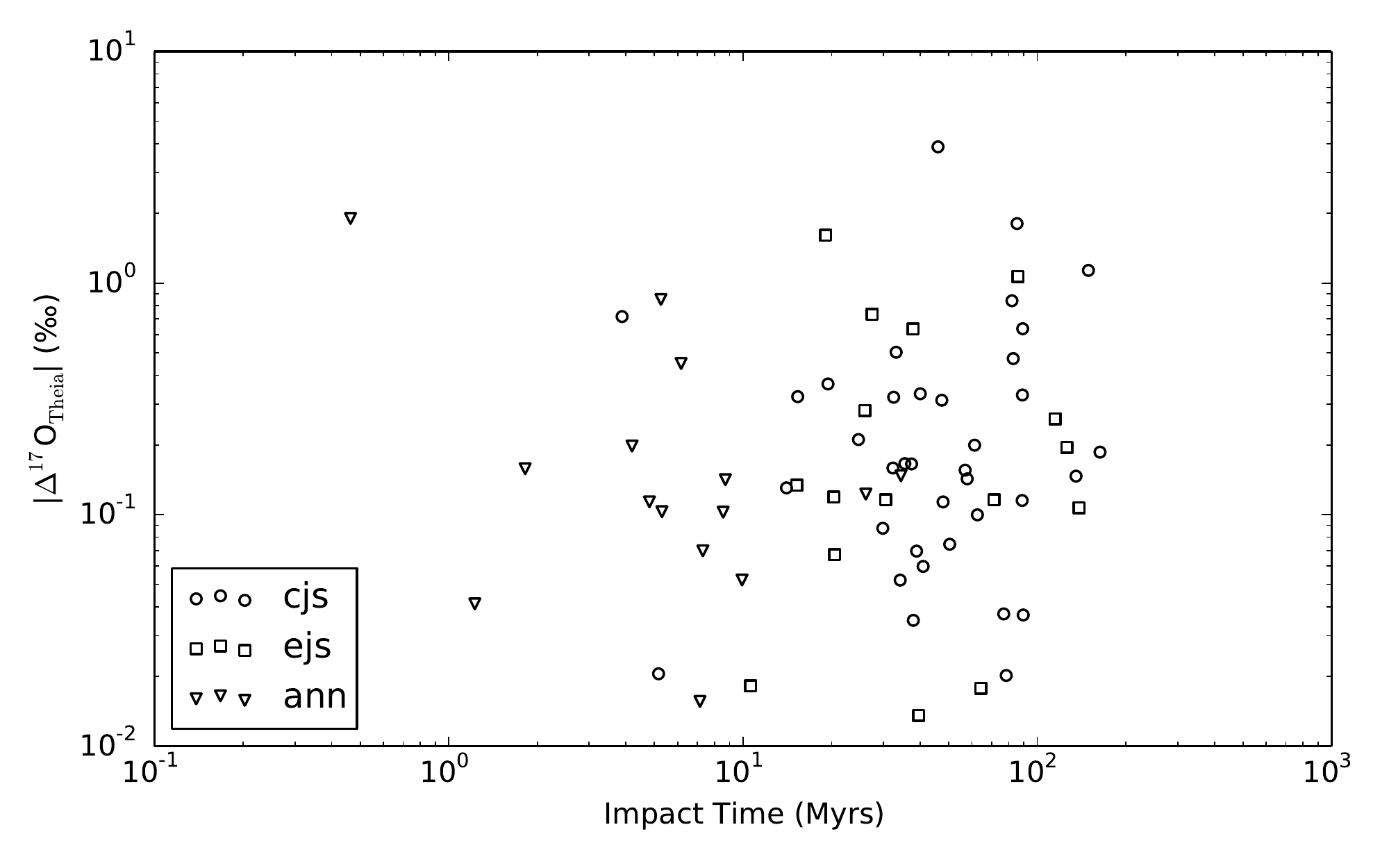}
\caption{The $|\Delta^{17}$O$|$ values of Theia analogs are plotted against the simulation time at which they collide with the Earth analog. Circles, squares, and triangles mark the CJS, EJS, and ANN simulation sets, respectively.}\label{fig:time}
\end{figure}

We can also search for a possible trend between $|\Delta^{17}$O$|$ and the time of the moon-forming impact. For instance, in the case of the CJS simulations, we may expect late-impacting Theia analogs to come exclusively from the distant outer part of the planet-forming disk, since accretion proceeds much more slowly in this region. If this were the case, then  late-colliding Theia analogs may show especially large differences between themselves and the Earth analogs. To search for such correlations, in Figure \ref{fig:time} we plot the $|\Delta^{17}$O$|$ value of each Theia analog against the time it collides with the Earth analog. As can be seen in the plot, there is no obvious correlation between impact time and how isotopically Earth-like a Theia analog is.

The results shown in Figure \ref{fig:theialine} do not necessarily indicate that Earth and Theia analogs have fundamentally different feeding zones. Even if the two bodies had the same feeding zone we would expect a large fraction of our Theia analogs to have different $\Delta^{17}$O values simply because they are smaller and sample the same pool of planetesimals fewer times. Thus, the population of Theia analogs should have a higher variance in $\Delta^{17}$O than the population of Earth analogs as long as the Theia analogs have lower masses.

\begin{figure}
\centering
\includegraphics[scale=.43]{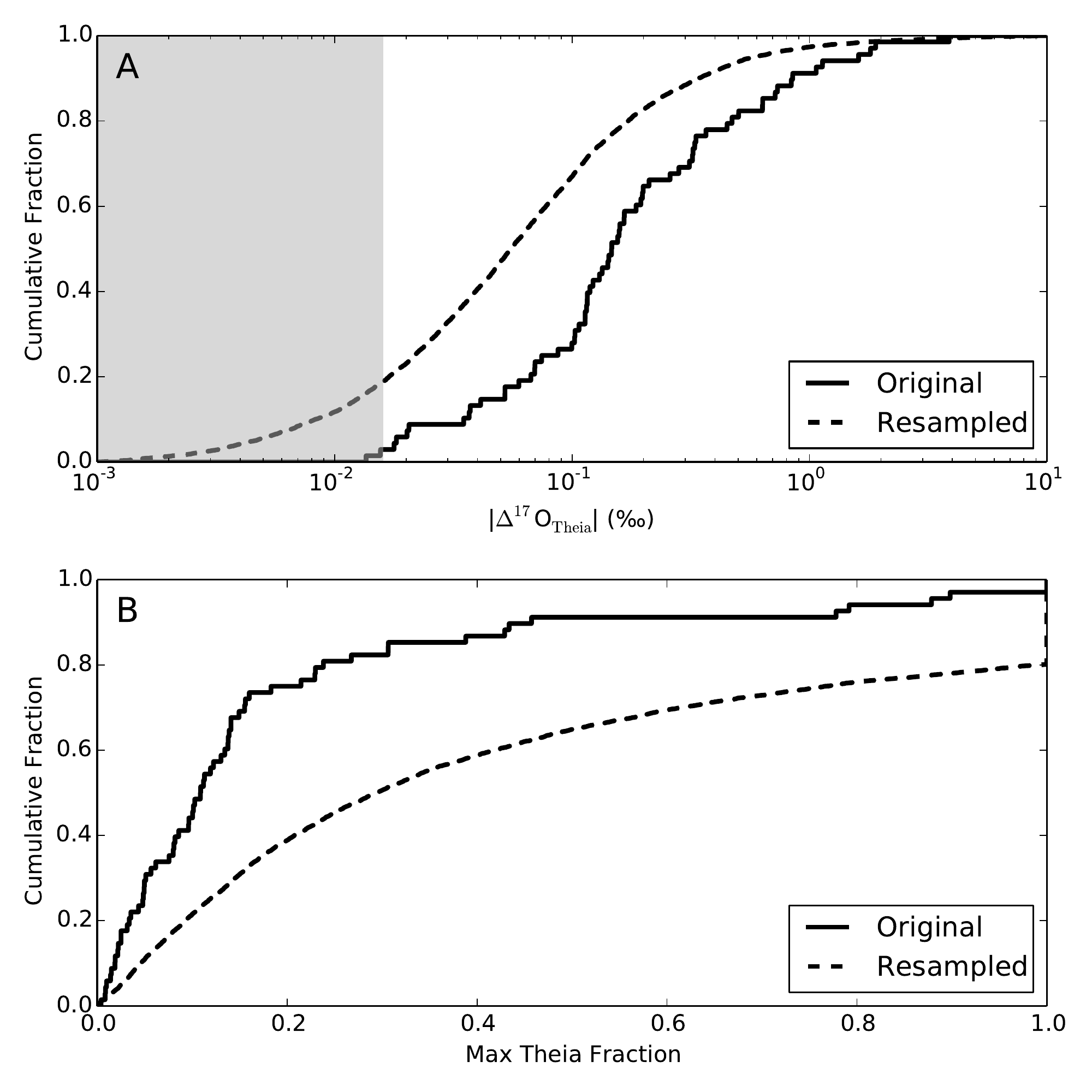}
\caption{{\bf A:} The $|\Delta^{17}$O$|$ values of Theia analogs in each simulation are shown. The dashed line shows the CDF for a set of artificial Theia analogs built by simply resampling the feeding zone of each Earth analog, while the solid line is the CDF for the actual Theia analogs constructed via accretion in our simulations. The shaded regions marks values of $|\Delta^{17}$O$|$ below 0.016\permil. {\bf B:} CDFs of the maximum contribution to the moon's mass that each Theia analog can make. Our artificial Theia analogs are shown with the dashed line and our actual Theia analogs constructed via accretion in our simulations are shown with the solid line.}\label{fig:theiafake}
\end{figure}

We can easily test whether Theia's lower mass is the dominant source of the large $\Delta^{17}$O values among Theia analogs. To do this, in each simulation we remove the Theia particles from our Earth analog, so we have an Earth feeding zone that includes only non-Theia particles. Next we redraw particles from our modified Earth feeding zone until we have drawn the equivalent mass of the original Theia analog that we just discarded. With this done, we have constructed an artificial Theia analog whose feeding zone is forced to be the same as the Earth analog. Using our modified Earth analog and our artificially constructed Theia analog, we can then predict a $\Delta^{17}$O for the artificially built Theia analog just as we did in Figure \ref{fig:theialine}. Moreover since we are constructing our new Theia analogs without dynamical accretion, we repeat this resampling process 100 times for each simulation to build up a smoother $\Delta^{17}$O distribution. The results of this process are shown in Figure \ref{fig:theiafake}A. Here we compare our artificially constructed Theia analogs with the real Theia analogs that were generated via accretion in our simulations. We see that the Theia analogs built from resampling Earth's feeding zone have significantly smaller $|\Delta^{17}$O$|$ values (median of 0.054\permil) than our original Theia analogs (median of 0.147\permil). In addition, only 2.9\% (2/68) of our original Theia analogs had $|\Delta^{17}$O$|<0.016$\permil, whereas 20\% of our artificially built Theia analogs have $|\Delta^{17}$O$|<0.016$\permil. We can also again look at the maximum contribution that our set of Theia analogs are allowed to make to the moon's composition in Figure \ref{fig:theiafake}B. We see that our artificially constructed Theia analogs can make up a significantly larger fraction of the moon's mass (median value of 29.4\%) compared to the original set of Theia analogs generated by our simulations (median value of 10.9\%). 

\begin{figure}
\centering
\includegraphics[scale=.43]{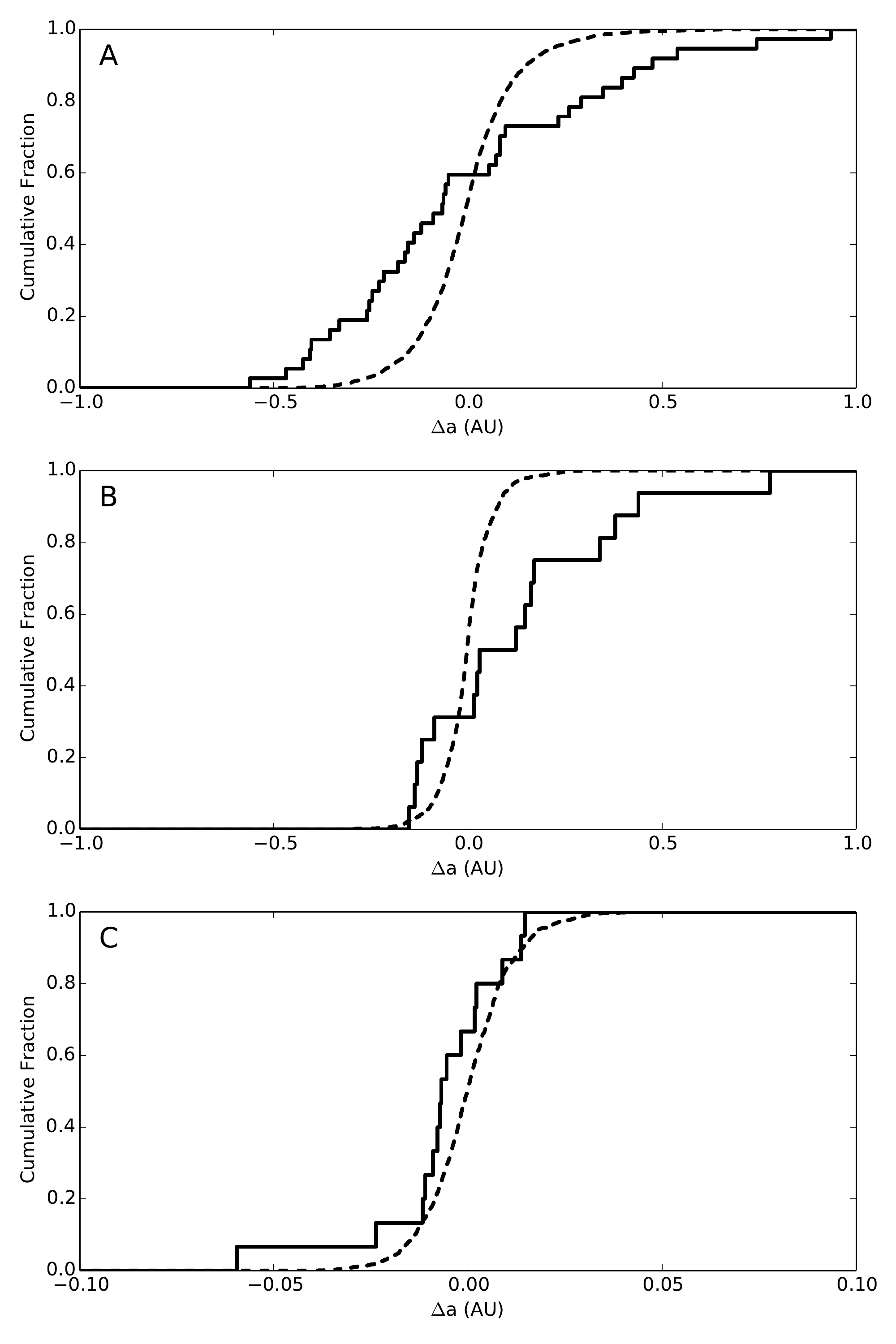}
\caption{The difference between the mean formation semimajor axis of each Earth and Theia analog is shown. (Negative $\Delta a$ implies that Theia formed from material that is interior to Earth's material.) The solid line is a CDF of the semimajor axis difference between Earth analogs and the Theia analogs formed in our simulations via accretion. The dashed line is a CDF for the Theia analogs that are artificially constructed by simply resampling the Earth's feeding zone. The results of the CJS, EJS and ANN simulations are shown in panels A, B, and C, respectively.}\label{fig:deltaa}
\end{figure}

Although Figure \ref{fig:theiafake} shows that some of the variance in the $\Delta^{17}$O values of Theia analogs is due to the bodies' smaller masses, it also demonstrates that Theia analogs tend to have fundamentally different feeding zones than Earth analogs, and the differences in feeding zone location between Earth and Theia is what drives most of the $\Delta^{17}$O difference between the Earth and Theia analogs in our simulations. This can be more clearly seen in Figures \ref{fig:deltaa}A--C. In these panels we calculate the difference between the mean formation semimajor axis of each Earth and Theia analog pair. A CDF of these differences is plotted for each simulation set. This is done for the actual Theia analogs built in the simulations as well as the artificially generated Theia analogs used to make Figure \ref{fig:theiafake}. If Theia and the Earth were derived from the same material the two CDFs would overlay each other in each panel. Instead, the real Theia analogs show a  significantly greater spread in formation semimajor axis. The plots indicate that although Theia is roughly equally likely to form from material interior to the Earth as it is to form from exterior material, its feeding zone is usually distinct from the rest of Earth.  Using a K-S test, we compare the artificial Theia CDF with the real Theia CDF for each simulation set. This comparison yields p-values of 0.13\%, 0.13\%, and 7.8\% for the CJS, EJS, and ANN simulations, respectively. 

\subsubsection{Step Function}

In contrast to smooth linear $\Delta^{17}$O distributions, we can also explore initial $\Delta^{17}$O distributions designed to give sharp contrasts between outer and inner terrestrial planets. To do this, we model $\Delta^{17}$O distributions with a step function. To build our step function, we first choose a contrast for the inner and outer values of $\Delta^{17}$O. Next, we set the heliocentric distance at which the step occurs in order to produce a difference in $|\Delta^{17}$O$|$ for Mars and Earth analogs that is as close as possible to 0.32\permil. (Because the smallest masses in our simulations are of order 10$^{-3}$ M$_{\oplus}$, it is not always possible to exactly match a 0.32\permil~ difference with a step function.) 

\begin{figure}
\centering
\includegraphics[scale=.43]{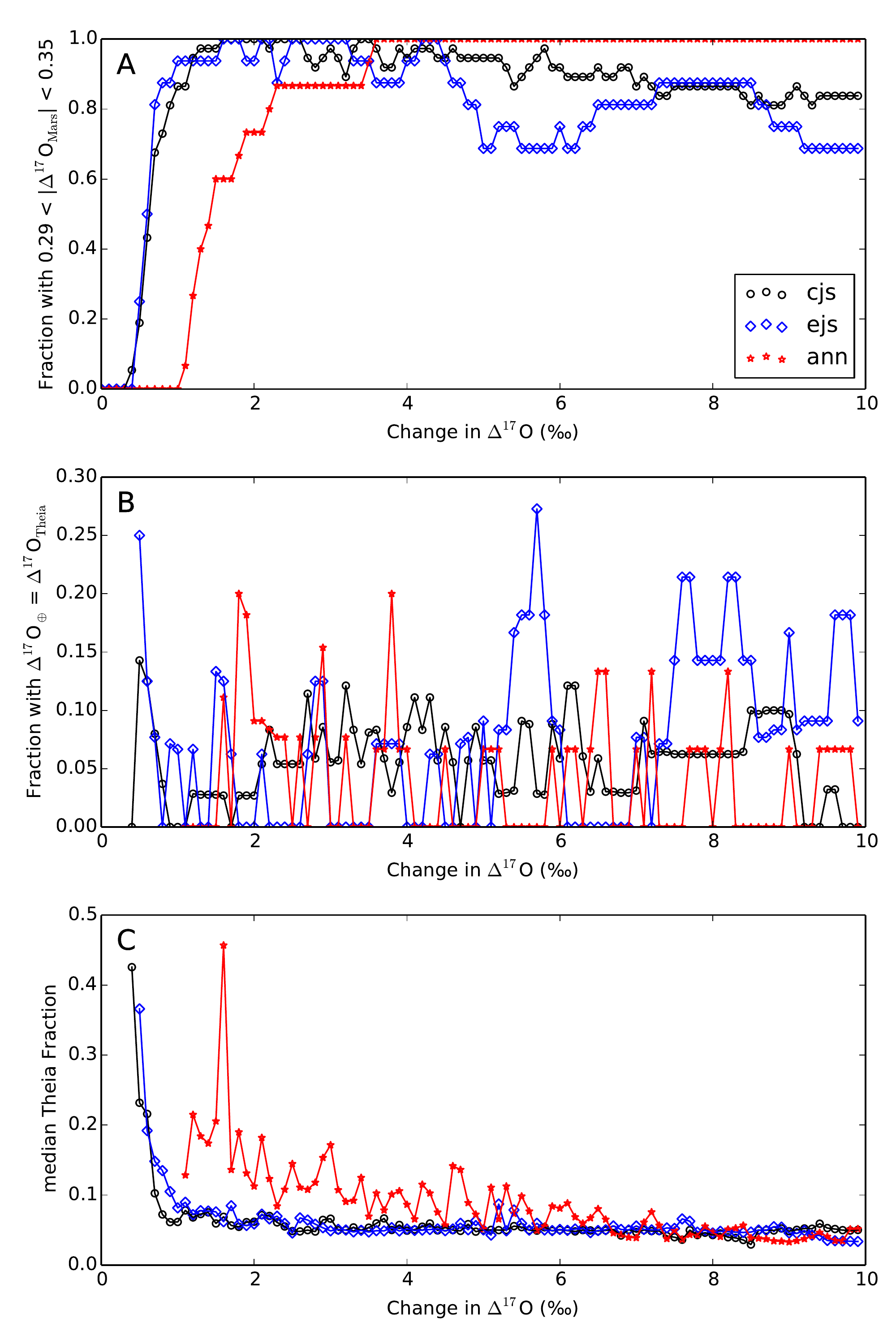}
\caption{{\bf A:} The fraction of Mars analogs that are able to satisfy $|\Delta^{17}$O$|=0.32\pm0.03$\permil~ (compared to the system's Earth analog) as a function of the chosen jump in $|\Delta^{17}$O$|$ in the initial $|\Delta^{17}$O$|$ step-function distribution (for all simulations containing Earth, Mars, and Theia analogs). {\bf B:} The fraction of systems that are found to have Earth and Theia analogs with $|\Delta^{17}$O$|$ within 0.016\permil~ of each other as a function of the chosen jump in $|\Delta^{17}$O$|$ in the initial $|\Delta^{17}$O$|$ step-function distribution. {\bf C:} Given the predicted $|\Delta^{17}$O$|$, the maximum allowable contribution of Theia to the Moon's mass is calculated. The median value of the maximum allowable Theia contribution is plotted as a function of the step size in in the assumed initial $|\Delta^{17}$O$|$ distribution. CJS, EJS, and ANN simulation results are shown with black circles, blue diamonds, and red stars, respectively, in all three panels.}\label{fig:theiastep}
\end{figure}

The results of this approach will clearly depend on the contrast we choose between the inner and outer values of $\Delta^{17}$O. For this reason, in each simulation we test a range of contrast values. Guided by the spread of mean $\Delta^{17}$O values seen among the chondritic meteorite groups \citep{burbineobrien04, ozima07}, we explore contrasts between 0 and 10\permil. In some cases, it is impossible to attain a difference near 0.32\permil ~for Earth and Mars analogs' values no matter what heliocentric distance the step occurs. We therefore only consider simulations that have attained a Martian $|\Delta^{17}$O$|$ of $0.32\pm0.03$ relative to Earth. In Figure \ref{fig:theiastep}A, we show the fraction of our simulations that have met this criterion as a function of the chosen contrast. One can see that a $|\Delta^{17}$O$|$ a difference of 0.32\permil~ between Earth and Mars cannot be attained for small step sizes. This is not surprising, as no difference between the two planets' compositions will occur if the step contrast is zero. For CJS and EJS simulations, the optimal contrast is between $\sim$1 and $\sim$4\permil. Beyond 4\permil~ the number of Earth and Mars analogs with $|\Delta^{17}$O$|$ near 0.32\permil~ falls off again. The reason for this is that while there is often a large amount of overlap between the feeding zones of Earth and Mars analogs in these simulations, some simulations yield fairly distinct feeding zones. As a result, it sometimes becomes impossible for the two analog planets to have a small $|\Delta^{17}$O$|$ value if a large contrast is chosen (no matter what the heliocentric distance of the step is). Alternatively, the ANN simulations require a larger contrast in the initial $|\Delta^{17}$O$|$ distribution to attain Earth and Mars analogs with $|\Delta^{17}$O$|$ near 0.32\permil, because planets forming from a narrow annulus have nearly entirely overlapping feeding zones.

For step function contrasts in which at least half of our simulations match the Martian $|\Delta^{17}$O$|$, we use this same step function to calculate the $|\Delta^{17}$O$|$ value for Theia analogs. Of the simulations that yield a Martian $|\Delta^{17}$O$|$ of $0.32\pm0.03$, we plot the fraction of Theia analogs that have $|\Delta^{17}$O$|<0.016$\permil~ in Figure \ref{fig:theiastep}B. From this plot we see that the $|\Delta^{17}$O$|$ values for Theia analogs are typically greater than 0.016\permil, regardless of the position and contrast of the step. Examining all the versions of the initial $|\Delta^{17}$O$|$ we employ, the mean fraction of Theia analogs with $|\Delta^{17}$O$|<0.016$\permil~ is 5.5\%, 7.2\%, and 4.4\% for our CJS, EJS, and ANN simulations respectively. These frequencies of Theia analogs with Earth-like oxygen signatures are comparable to those found when we assumed a linear distribution of $\Delta^{17}$O. Thus, an Earth-like composition for Theia is an unlikely outcome of terrestrial planet formation for either family of initial $\Delta^{17}$O distributions.

Once again, we also determine how much material each Theia analog can contribute to the Moon's composition. We calculate the median value of this allowable contribution for each step function contrast we employ for the initial $\Delta^{17}$O distribution. The results of this are shown in Figure \ref{fig:theiastep}C. Similar to our linear $\Delta^{17}$O distributions, we find that typically Theia material must not account for more than 5--15\% of the Moon's composition. The exception to this occurs at small step function contrasts, where Theia is sometimes allowed to contribute up to 40\% of the Moon's mass. However, as can be seen in Figure \ref{fig:theiastep}A, these small step contrasts have difficulty consistently reproducing Mars' isotopic difference.

\subsubsection{Random Distribution}

The fact that the $\Delta^{17}$O of Vesta is lower than Earth, while that of Mars is higher is suggestive that the initial distribution of $\Delta^{17}$O did not monotonically increase or decrease with heliocentric distance. This would invalidate both linear trends and step functions for the initial radial $\Delta^{17}$O distribution. The opposite extreme is to assume that there is no correlation between initial $\Delta^{17}$O values and initial semi-major axis. To perform this exercise, we generate a $\Delta^{17}$O value for each initial particle in our simulations by randomly drawing from a Gaussian distribution centered on $\Delta^{17}$O $ = 0$. By combining the mass-weighted $\Delta^{17}$O values of our original particles we then calculate the final $|\Delta^{17}$O$|$ of Earth, Mars, and Theia analogs. We rerun this random generation of $\Delta^{17}$O 10,000 times on each ensemble (CJS, EJS, and ANN) to build up our statistics. For each ensemble, the standard deviation of our initial $\Delta^{17}$O values is adjusted so that the median absolute difference between the $\Delta^{17}$O of Earth and Mars analogs is 0.32\permil. To meet this requirement, the standard deviations of our initial $|\Delta^{17}$O$|$ distributions are set to 5.80, 4.86, and 2.82\permil~ for the CJS, EJS, and ANN simulation sets, respectively.

\begin{figure}
\centering
\includegraphics[scale=.43]{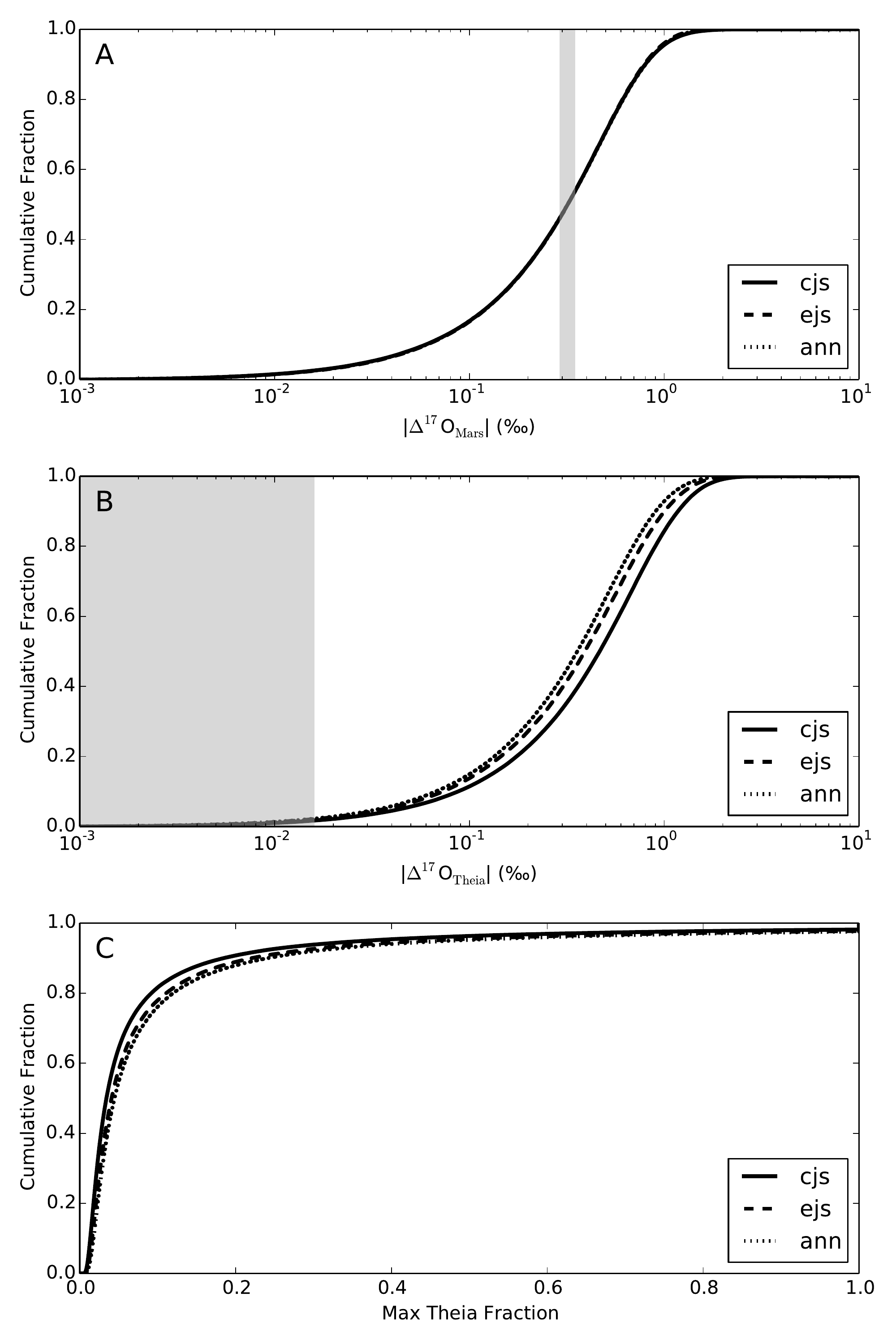}
\caption{{\bf A:} The distribution of $|\Delta^{17}$O$|$ for Mars analogs if we assume that the initial $\Delta^{17}$O values of planetesimals are randomly drawn from a Gaussian distribution (for simulations that contain Earth, Mars, and Theia analogs). The shaded region marks 0.29\permil $ < |\Delta^{17}$O$| < 0.32 \permil$. Initial $\Delta^{17}$O values of embryos and planetesimals were redrawn 10000 times to build our statistics. {\bf B:} The distribution of $|\Delta^{17}$O$|$ values for Theia analogs for the same simulations as in panel A. The shaded region marks $|\Delta^{17}$O$| < 0.016 \permil$. {\bf C:}Given the Theia $|\Delta^{17}$O$|$ values in panel B, the maximum allowable contribution to our Moon's mass is calculated for each Theia analog. The CDFs of these allowable contributions are shown here. CJS, EJS, and ANN simulation results are displayed with the solid, dashed, and dotted CDFs, respectively, in all three panels.}\label{fig:theianorm}
\end{figure}

In Figure \ref{fig:theianorm}A, we display the CDF of $|\Delta^{17}$O$|$ values for our Mars analogs for the 10,000 random oxygen isotope generations in CJS, EJS, and ANN. As can be seen, the $|\Delta^{17}$O$|$ of Mars analogs varies substantially between $\sim$0.01 and $\sim$1\permil~ with the median near 0.32\permil. In Figure \ref{fig:theianorm}B, we plot the CDFs of the $|\Delta^{17}$O$|$ values for Theia analogs in these same simulations. In all three sets of our simulations, there is only a tiny probability ($\sim$2\% in each simulation set) that a Theia analog will have a $|\Delta^{17}$O$|$ below 0.016\permil. The level of variance in $\Delta^{17}$O required to generate the observed oxygen difference between Earth and Mars analogs almost always also yields Theia analogs that differ substantially from Earth. Thus, once again we find it quite unlikely that an Earth-like composition for Theia is produced via terrestrial planet accretion.

Figure \ref{fig:theianorm}C shows the maximum contribution that each Theia analog can make to the Moon's mass. Again, only small mass fractions are typically allowed. The median values of the allowed Theia contribution to the Moon's mass are 3.4\%, 4.1\%, and 4.5\% for the CJS, EJS, and ANN simulations, respectively.

In our previous hypothetical $\Delta^{17}$O distributions with an explicit dependence on heliocentric distance, the final $\Delta^{17}$O values of the planets and impactors are greatly affected by the level of radial mixing within the protoplanetary disk during terrestrial planet formation. However, when one assumes no dependence on heliocentric distance (as is the case here) the final compositions of planets and their impactors are largely independent of the simulation's dynamics. Instead, the expected deviation from the disk-averaged $\Delta^{17}$O value for each planet and impactor is a function of the number of planetesimals and embryos that the body accreted during the simulation; i.e., its mass. This is just a manifestation of the Central Limit Theorem, and since we tune our initial $\Delta^{17}$O to typically yield substantial differences between Earth and Mars, it is no surprise that Theia analogs (which tend to be less massive bodies) usually have $\Delta^{17}$O values that are at least as large as the simulations' Mars analogs.

\subsubsection{Bimodal $\Delta^{17}$O Values}\label{sec:bimod}

One last scenario that we explore for the initial $\Delta^{17}$O values in the protoplanetary disk is one where all embryos have one fixed $\Delta^{17}$O value and all planetesimals have another. One could imagine such a situation arising if bodies' $\Delta^{17}$O values are affected by their mean formation time within the solar nebula or different parent body processing. In this case, each protoplanet's $\Delta^{17}$O value would be set by the fraction of its mass contributed by planetesimals vs embryos. (Note that the ANN simulations are excluded from this analysis since they began with only embryos.)

\begin{figure}
\centering
\includegraphics[scale=.43]{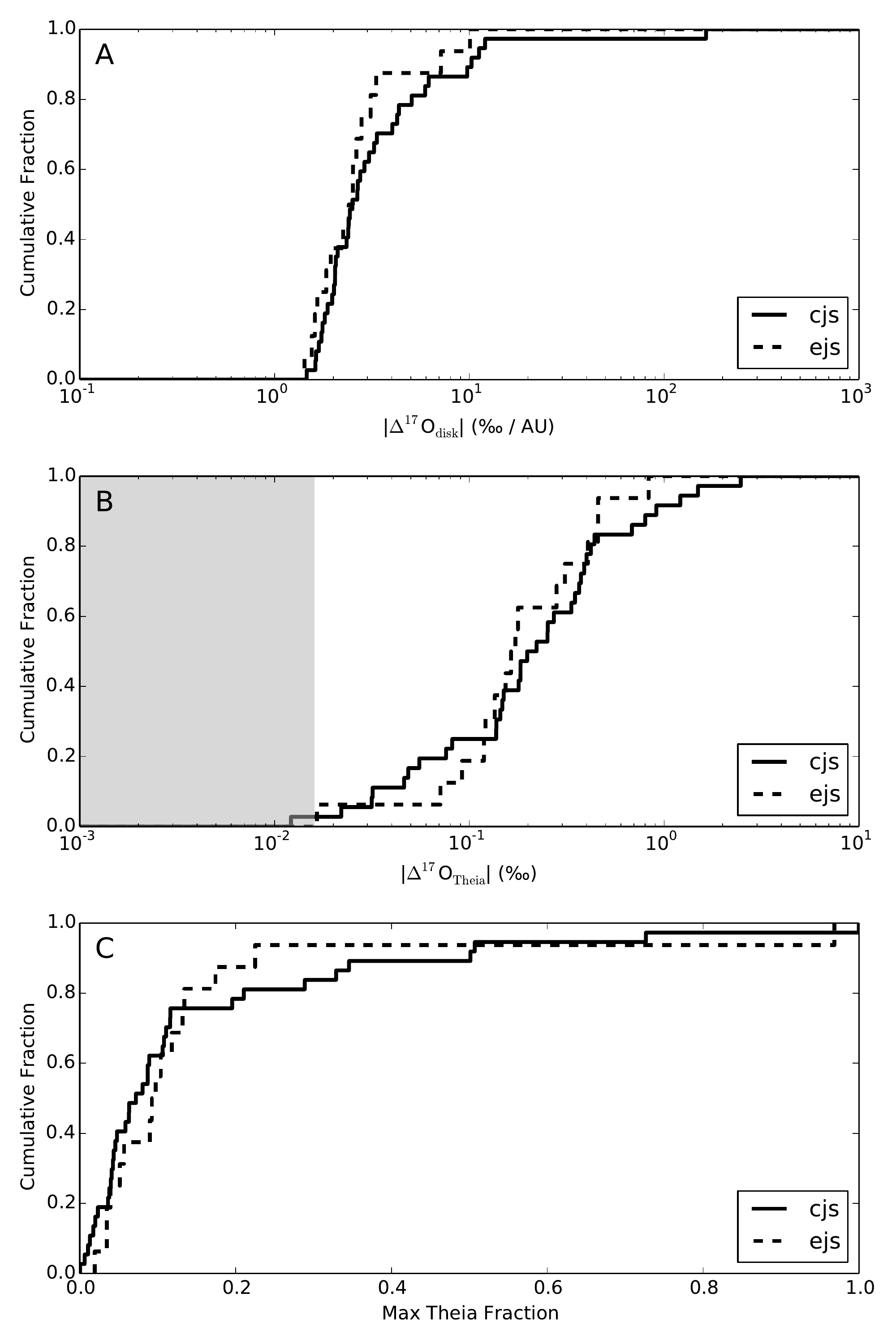}
\caption{{\bf A:} Assuming two different initial $\Delta^{17}$O for planetesimals and embryos, a CDF of the $|\Delta^{17}$O$|$ differences between planetesimals and embryos required to reproduce the observed Martian isotopic value is shown for all of our simulations that form Earth, Mars, and Theia analogs. {\bf B:} Using the distributions specified in panel A, the $|\Delta^{17}$O$|$ values of Theia analogs in each simulation are calculated and shown here. The shaded regions marks values of $|\Delta^{17}$O$|$ below 0.016\permil. {\bf C:} Given the Theia $|\Delta^{17}$O$|$ values in panel B, the maximum allowable contribution to our Moon's mass is calculated for each Theia analog. The CDFs of these allowable contributions are shown here. CJS, EJS, and ANN simulation results are displayed with the solid, dashed, and dotted CDFs, respectively, in all three panels.}\label{fig:theiaplem}
\end{figure}

To set the $\Delta^{17}$O difference between planetesimals and embryos in each simulation, we simply calculate the planetesimal fractions of the Earth and Mars analogs and then fix the $\Delta^{17}$O of planetesimals to yield a Martian $|\Delta^{17}$O$|$ of 0.32\permil. The distribution of $\Delta^{17}$O differences between planetesimals and embryos in our simulations is shown in Figure \ref{fig:theiaplem}A. We see that planetesimals typically must have $|\Delta^{17}$O$|$ values that are $\sim$2\permil~ different from the embryos in order to match the observed isotopic difference between Earth and Mars.

With the $\Delta^{17}$O of our initial embryos and planetesimals tuned, we can now determine the $|\Delta^{17}$O$|$ value predicted for Theia in each simulation. These are shown in Figure \ref{fig:theiaplem}B. We see that there are nearly no examples of Theia analogs with Earth-like oxygen isotopes, and the typical isotopic difference between Earth and Theia is 0.1--0.2\permil. Thus, if the Moon's composition is dominated by material from Theia, our simulations predict that its observed isotopic composition is a very improbable event.

This fact is emphasized in Figure \ref{fig:theiaplem}C where we calculate the maximum mass fraction that each Theia analog is allowed to contribute to the Moon's mass. We find that $\sim$90\% of our Theia analogs are prohibited from constituting more than $\sim$30\% of the Moon's mass, and the median allowable mass fractions are 7.2\% and 9.5\% for the CJS and EJS simulations sets, respectively. 

However, all of our simulations fix the initial mass ratio of planetesimals to embryos to be 1:1. \citet{jacmorb14} investigated planet formation with various embryo-to-planetesimal ratios and concluded that the total mass was likely dominated by embryos at the start of the giant impact phase of planet formation.  Furthermore, given the shorter collision timescales in the inner protoplanetary disk, it's likely that the true ratio of planetesimals to embryos varied with distance, and presumably the number of planetesimals would be greatest in the outer part of the disk. If this is the case, one could imagine Mars accreting many more planetesimals than the Earth and Theia. If so, an isotopically identical Earth and Theia may not conflict with an isotopically distinct Mars in this case. This could also potentially allow for an isotopically uniform inner solar nebula \citep{jav10,dauph14}, with Mars being polluted by more distant planetesimals. However, this scenario is untested and should be the subject of future studies.

\subsection{Potential Solutions: Massive Theias and High Velocity Impacts}\label{sec:theiamass}

In the previous section, we found that regardless of the initial $\Delta^{17}$O we impose on embryos and planetesimals, it is very unlikely for Theia analogs to be isotopically similar to Earth analogs. This presents a significant issue for the canonical moon formation scenario involving a Mars-mass impactor, since it generates a moon primarily composed of the impactor rather than the Earth. However, \citet{canup12} suggests that this issue can be resolved if Theia had a mass comparable to that of the proto-Earth. In this case, both the Earth and moon-forming disk are a roughly even mixture of the proto-Earth and Theia. (This scenario relies on the angular momentum of the Earth-Moon system later decreasing via an evection resonance with the Sun \citep{cukstew12}.)

\begin{figure}
\centering
\includegraphics[scale=.43]{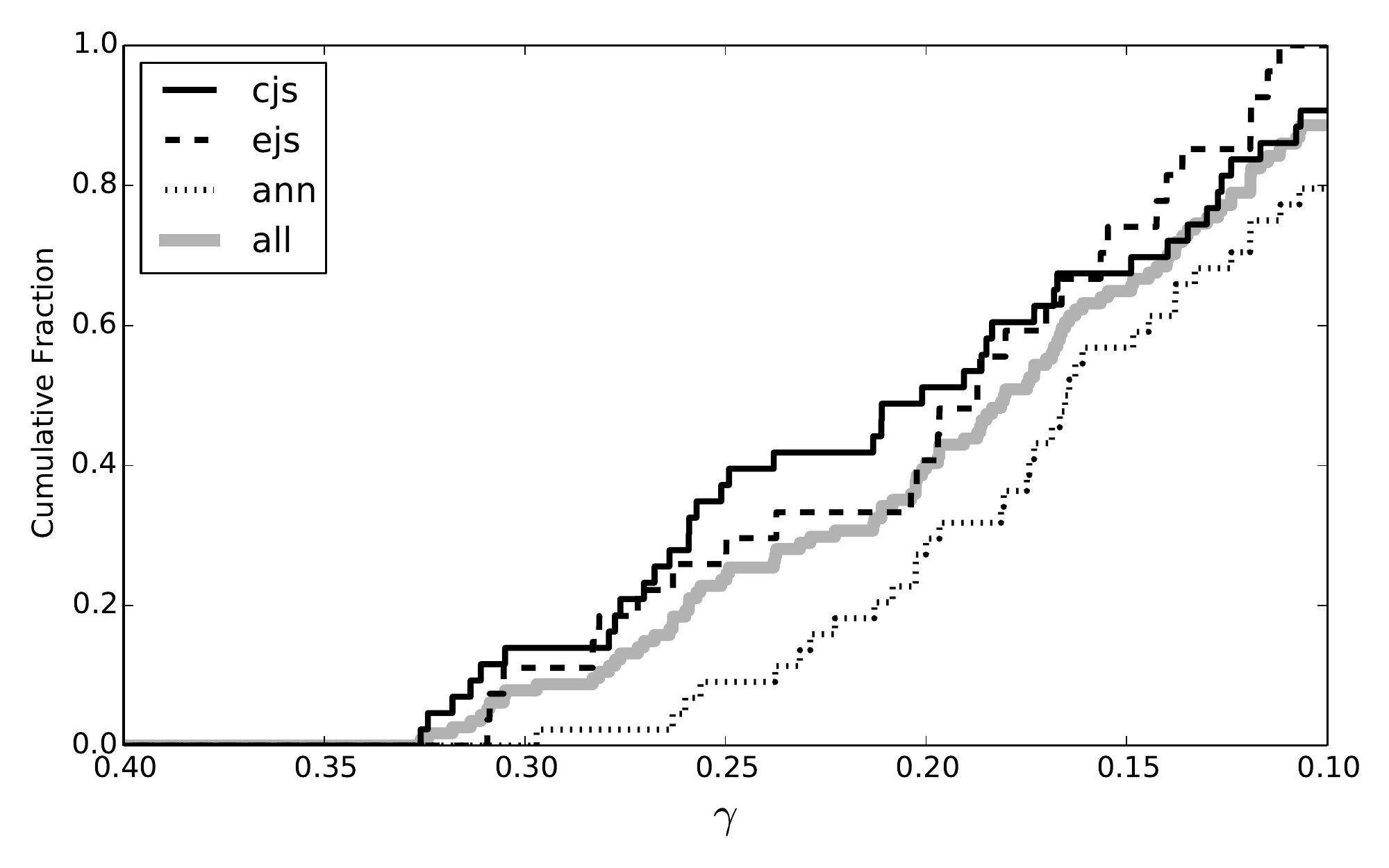}
\caption{The distribution of $\gamma$ values seen among Earth analogs in our simulations ($\gamma$ being the ratio of Theia's mass to the combined mass of Theia and the proto-Earth at the time of impact). Results from the CJS, EJS, and ANN simulations are shown with the black solid, dashed, and dotted CDFs, respectively. The combined CDF for all Earth analogs is marked with the thicker gray line.}\label{fig:theiamass}
\end{figure}

With our large number of terrestrial planet formation simulations, we can estimate the statistical likelihood that Theia's mass was comparable to the proto-Earth. To do this, we simply look at the distribution of mass ratios for Earth analogs struck by Theia analogs in our simulations. This distribution is shown in Figure \ref{fig:theiamass}, where we plot the parameter $\gamma$, which is the ratio of Theia's mass to the combined mass of Theia and the proto-Earth at the time of impact. To mix the Earth and Moon evenly enough, \citet{canup12} finds that Theia must have had $\gamma \gtrsim 0.4$. In Figure \ref{fig:theiamass}, we see that such collisions are not found in any of our simulations. Out of the 104 Earth analogs generated in our collisions, the largest recorded $\gamma$ is 0.325, and only 8.7\% of our Earth analogs experienced impacts with $\gamma>0.3$. Late impacts with $\gamma \gtrsim 0.4$ must be exceedingly rare, implying that a comparably massed Theia and proto-Earth is a very unlikely event. This result agrees with \citet{jacmorb14}, who also found that major mergers between protoplanets with similar masses are rare.

\begin{figure}
\centering
\includegraphics[scale=.43]{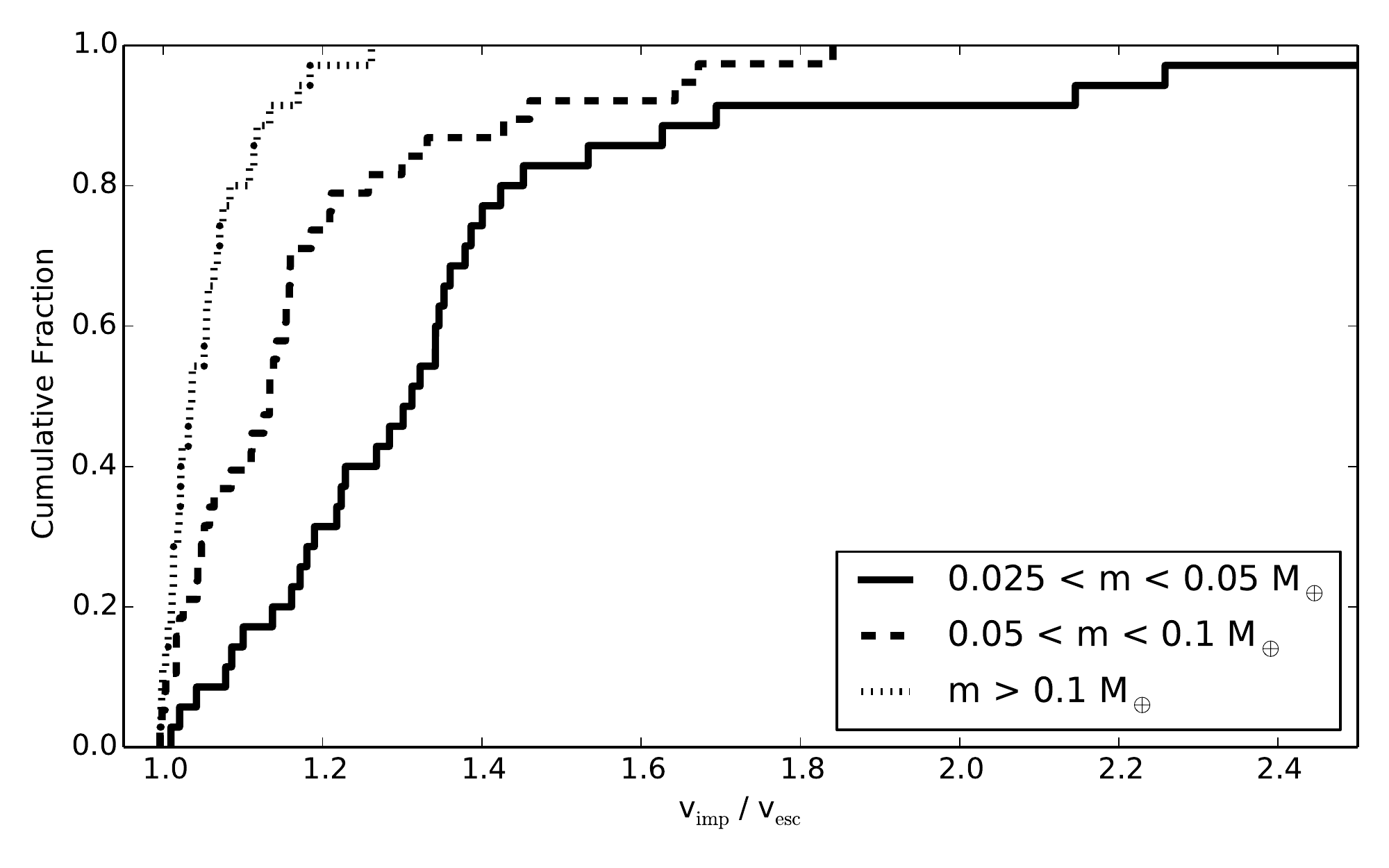}
\caption{The cumulative distribution of impact velocities between Earth and Theia analogs in the ANN simulations. Theia analogs are split into three different mass bins: $m$ = 0.025--0.05 M$_{\oplus}$ (solid line), $m$ = 0.05--0.1 M$_{\oplus}$ (dashed line), and $m>$ 0.1 M$_{\oplus}$ (dotted line). Impact velocity is calculated in terms of the mutual escape velocity of the Earth and Theia analogs.}\label{fig:theiavel}
\end{figure}

Instead of comparable masses for Theia and the proto-Earth, both \citet{reufer12} and \citet{cukstew12} invoke a higher impact velocity between Theia and the proto-Earth to produce an Earth-like moon. Unfortunately, our CJS and EJS simulations did not record the collisional velocities between embryos, so the impact velocity distribution in these simulation sets is not known. However, we do have the collisional velocity data from the ANN simulations, which had 35 impacts between Earth and Theia analogs. The cumulative distribution of impact velocities between Theia analogs and Earth analogs in ANN simulations is shown in Figure \ref{fig:theiavel}. In this plot we see that the median impact velocity is just a few percent greater than the mutual escape velocity of the proto-Earth and Theia analogs with $m > 0.1$ M$_{\oplus}$. Moreover, the largest impact velocity seen in our simulations is 126\% of the mutual escape velocity. For comparison, most of the successful collisions in \citet{cukstew12} require an impact velocity at least 150\% of the mutual escape speed. Thus, our ANN simulations suggest that such high velocity impacts are rare for Theia analogs with $m >$ 0.1 M$_{\oplus}$.

\citet{cukstew12} also explore impact scenarios involving Theia masses of 0.05 M$_{\oplus}$ and 0.025 M$_{\oplus}$ and find that such collisions can successfully produce an Earth-heavy composition for the Moon if the proto-Earth was spinning very rapidly before impact. Because of this finding, we also look at our collision statistics for last major mergers on Earth analogs that involve impacting bodies with masses below 0.1 M$_{\oplus}$. These are also shown in Figure \ref{fig:theiavel}. We see that smaller impactors do collide with the Earth at higher velocities, but the large majority are still below 150\% of escape velocity, the minimum velocity preferred in \citet{cukstew12}. Of the Theia analogs with 0.025 M$_{\oplus} < m < 0.05 $M$_{\oplus}$, 17\% strike the proto-Earth with a velocity that is more than 150\% of the mutual escape velocity. For Theia analogs with 0.05 M$_{\oplus} < m < 0.1 $M$_{\oplus}$, the percentage of high-velocity collisions falls to 7.9\%. Because the CJS and EJS simulations involve collisions of bodies spanning a larger range of semimajor axes than the ANN set, we may expect larger fractions of energetic collisions in these two simulation sets. Unfortunately, these fractions are not known. However, in order for the \citet{cukstew12} mechanism to succeed, the Earth's spin rate also has to be exceptionally high (a period of $\sim$2.3 hours). Such a high spin rate is likely achieved with a very large merger event before the Moon-forming impact, and we have already shown in Figure \ref{fig:theiamass} that massive impactors are rare. Thus, production of our moon with a high velocity impact may also be a relatively low probability event.

\subsection{Comparison to Venus Analogs}\label{sec:venuscomp}

With four different families of initial $\Delta^{17}$O distributions set by the accretion histories of Earth and Mars analogs, we can now predict the $\Delta^{17}$O values that these would yield for Venus by examining the accretion histories of Venus analogs in our simulations. Because we need Earth and Mars analogs to define our initial $\Delta^{17}$O distributions, we only use simulations that form analogs for Venus, Earth, and Mars. (This set of 70 simulations is slightly different from the last section since we drop the requirement for Theia analogs and replace it with Venus analogs.) As in the previous section we examine the outcomes of four different initial $\Delta^{17}$O distributions for our embryos and planetesimals: a linear distribution, a step-function distribution, a random distribution, and two different values for planetesimals and embryos.

\begin{figure}
\centering
\includegraphics[scale=.43]{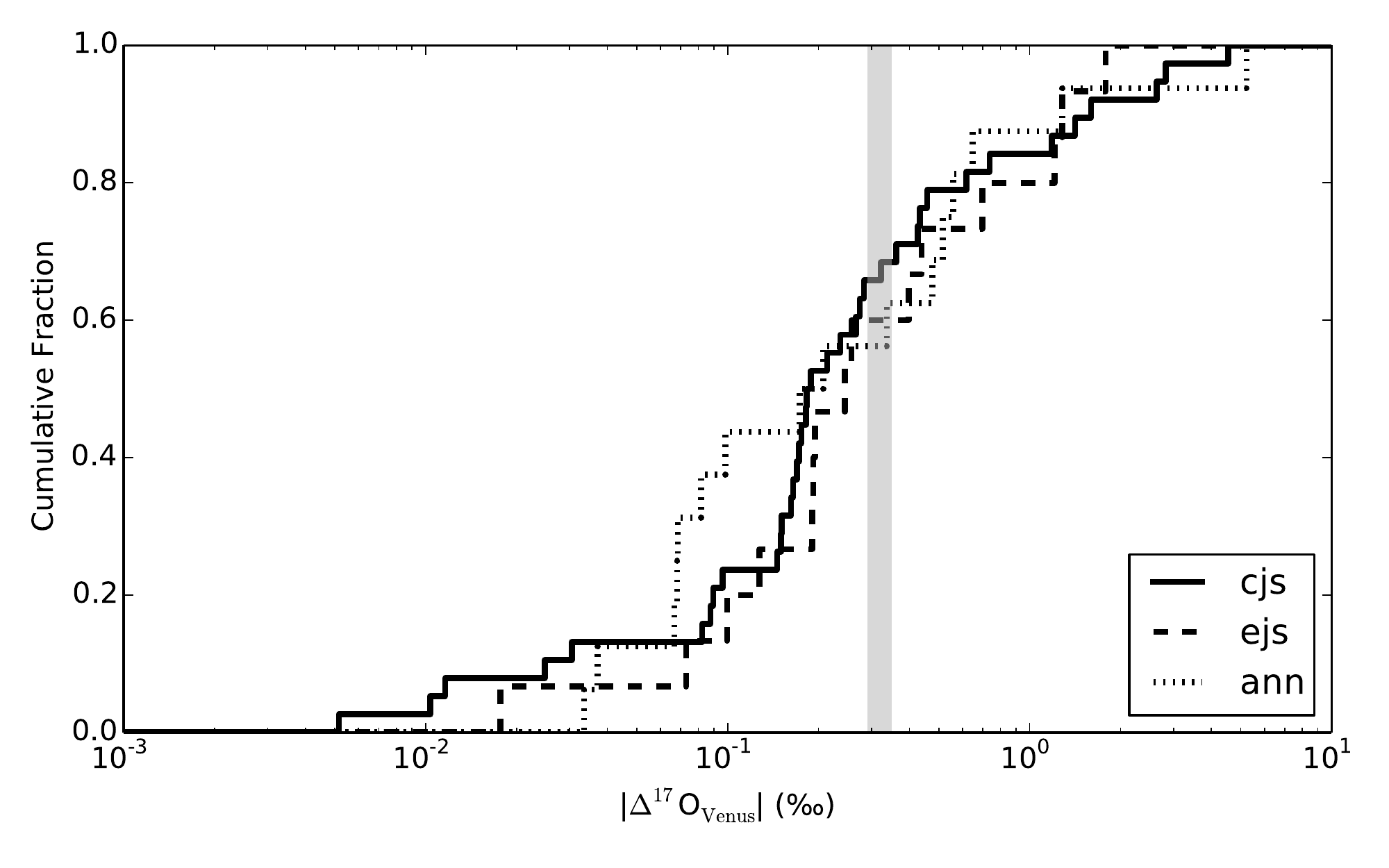}
\caption{The distribution of $|\Delta^{17}$O$|$ predicted for Venus analogs if we assume that the initial distribution of $\Delta^{17}$O varies linearly with heliocentric distance. The predicted values are shown for the CJS, EJS, and ANN simulations with the solid, dashed, and dotted CDFs, respectively. The shaded region marks 0.29\permil $ < |\Delta^{17}$O$| < 0.32 \permil$, the approximate observed value for Mars. }\label{fig:venusline}
\end{figure}

In Figure \ref{fig:venusline}, we show the distributions of $|\Delta^{17}$O$|$ found for our Venus analogs if we assume the initial $\Delta^{17}$O among embryos and planetesimals varies linearly with their heliocentric distance. As before, we force the linear distributions to yield $|\Delta^{17}$O$| \simeq0.32$\permil~ for the Mars analog in each simulation. When this is done, we find that the Venus analogs are unlikely to have similar oxygen isotope compositions to the Earth. In fact for each of our three simulation sets (CJS, EJS, and ANN), the median value of $|\Delta^{17}$O$|$ is only slightly more Earth-like than Mars. This is indicative of the unique feeding zones that Earth and Venus analogs have in a given system. Because Earth and Venus analogs have higher minimum masses than Mars analogs, it was suspected that the greater number of accretion events may cause the two planets to closely converge to the same average $\Delta^{17}$O value. However, this is not the case.

\begin{figure}
\centering
\includegraphics[scale=.43]{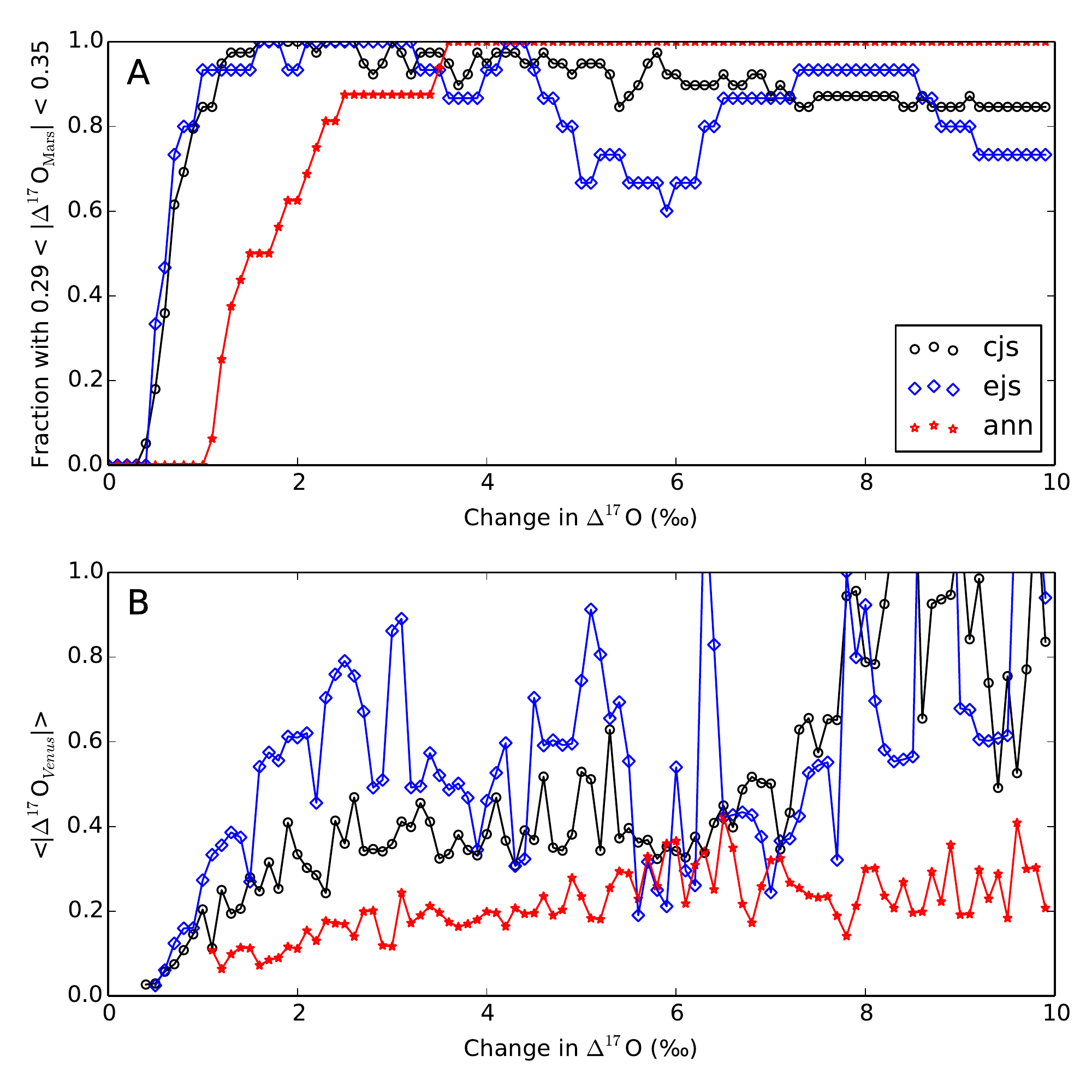}
\caption{{\bf A:} The fraction of Mars analogs that are able to satisfy $|\Delta^{17}$O$|=0.32\pm0.03$\permil~ (compared to the system's Earth analog) as a function of the chosen jump in $|\Delta^{17}$O$|$ in the initial $|\Delta^{17}$O$|$ step-function distribution (for all simulations containing Venus, Earth, and Mars analogs). CJS, EJS, and ANN simulation results are shown with black circles, blue diamonds, and red stars, respectively. {\bf B:} The median $|\Delta^{17}$O$|$ value predicted for Venus analogs as a function of the chosen jump in $|\Delta^{17}$O$|$ imposed for the initial $|\Delta^{17}$O$|$ step-function distribution. CJS, EJS, and ANN simulation results are shown with black circles, blue diamonds, and red stars, respectively.}\label{fig:venusstep}
\end{figure}

We now repeat this exercise, but instead the initial $\Delta^{17}$O distribution is modeled with a step function with a given contrast and heliocentric position to yield 0.29\permil~$< |\Delta^{17}$O$| < $0.35\permil~ for the Mars analog of each system. For each chosen step function contrast, we calculate the median value of $|\Delta^{17}$O$|$ for our Venus analogs in each ensemble (CJS, EJS, and ANN) if at least half of the simulations yielded a Martian $|\Delta^{17}$O$|$ between 0.29\permil~ and 0.35\permil. These median values are displayed in Figure \ref{fig:venusstep}B. As can be seen, the expected $|\Delta^{17}$O$|$ for Venus analogs is always at least 0.1\permil~ different from Earth (which by definition is 0) and are often equal or greater than than Martian  $|\Delta^{17}$O$|$ to which the initial distribution is tuned. It is also clear that the ANN simulations consistently yield the most Earth-like oxygen compositions for Venus. This is presumably because the ANN simulations more often yield Venus and Earth analogs that are much more massive than the Martian analogs \citep{hans09}, and they more uniformly sample the narrow annulus of embryos and planetesimals compared to CJS and EJS simulations.

\begin{figure}
\centering
\includegraphics[scale=.43]{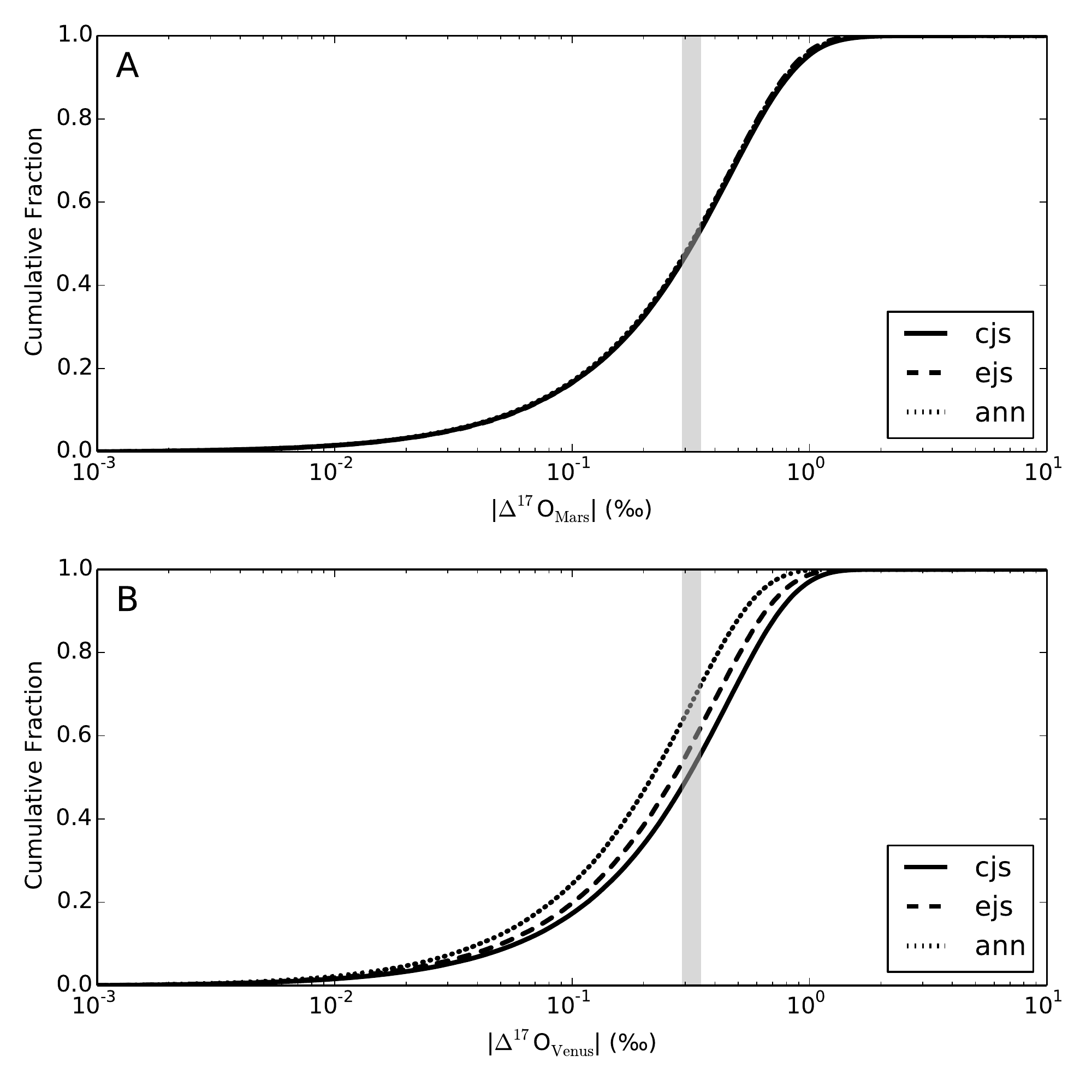}
\caption{{\bf A:} The distribution of $|\Delta^{17}$O$|$ for Mars analogs if we assume that the initial $\Delta^{17}$O values of planetesimals are randomly drawn from a Gaussian distribution (for simulations that contain Earth, Mars, and Venus analogs). CJS, EJS and ANN simulations are shown with solid, dashed, and dotted CDFs, respectively. The shaded region marks 0.29\permil $ < |\Delta^{17}$O$| < 0.32 \permil$. Initial $\Delta^{17}$O values of embryos and planetesimals were redrawn 10000 times to build our statistics. {\bf B:} The distribution of $|\Delta^{17}$O$|$ values for Venus analogs for the same simulations as in panel A. The shaded region again marks 0.29\permil $ < |\Delta^{17}$O$| < 0.32 \permil$.}\label{fig:venusnorm}
\end{figure}

Next we test the assumption that the initial $\Delta^{17}$O of each embryo and planetesimal is uncorrelated with heliocentric distance, and we draw them from a Gaussian distribution designed to typically yield a Martian $|\Delta^{17}$O$|$ near 0.32\permil. This is repeated 10,000 times on each simulation to build up robust statistics. Again, Figure \ref{fig:venusnorm}B shows that the $|\Delta^{17}$O$|$ values for Venus analogs are distinctly un-Earthlike. In fact, they once again often take values similar to Mars. 

\begin{figure}
\centering
\includegraphics[scale=.43]{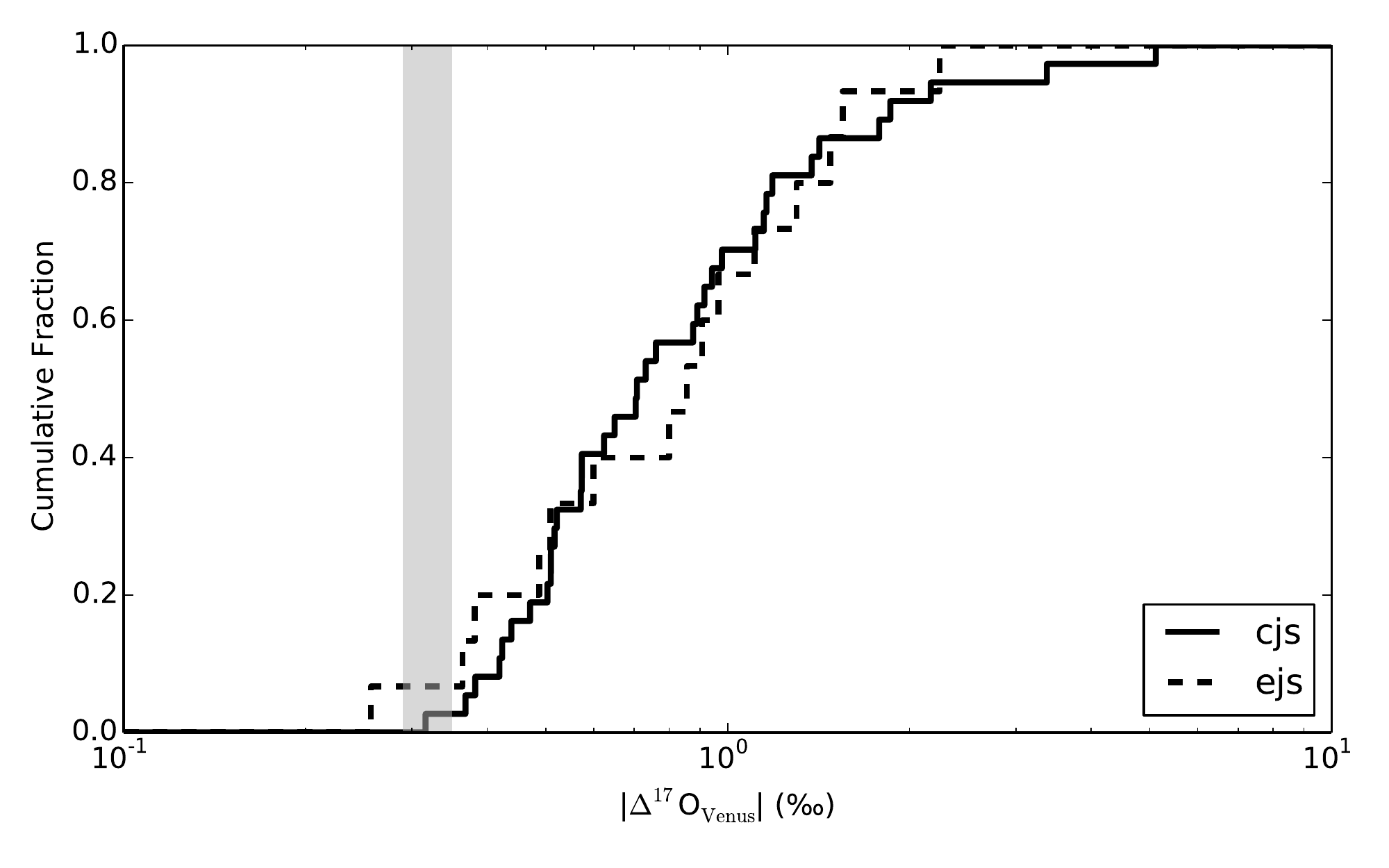}
\caption{The distribution of $|\Delta^{17}$O$|$ predicted for Venus analogs if we assume two different values of $\Delta^{17}$O for planetesimals and embryos. The predicted values are shown for the CJS, EJS, and ANN simulations with the solid, dashed, and dotted CDFs, respectively. The shaded region marks 0.29\permil $ < |\Delta^{17}$O$| < 0.32 \permil$, the approximate observed value for Mars. }\label{fig:venusplem}
\end{figure}

Finally, we again explore the possibility that planetesimals are born with one fixed $\Delta^{17}$O and embryos are born with another. As in section \ref{sec:bimod}, we set the two different values to yield $\Delta^{17}$O = 0.32\permil~ for the Mars analog in each simulation (relative to the Earth analog). Using this initial $\Delta^{17}$O distribution we predict the $\Delta^{17}$O values for the Venus analogs of our simulations. These are shown in Figure \ref{fig:venusplem}. We see in this figure that this analysis predicts that Venus should be even more isotopically different from the Earth than Mars. The median $|\Delta^{17}$O$|$ values for Venus are 0.73\permil~ and 0.86\permil~ for the CJS and EJS simulations, respectively. The reason for this is that the planetesimal mass fraction is almost always highest for the Earth, second highest for Mars, and third highest for Venus analogs in our simulations. Thus, Venus is almost always predicted to be more dissimilar from Earth than Mars. 

Although Venus is nearly the same mass as Earth, we find that it is likely to have a significantly different $\Delta^{17}$O value than the Earth. Thus, we find that Earth-size (and Venus-size) planets do not usually gravitate toward the same disk-averaged $\Delta^{17}$O. This finding is unchanged for each of our four different initial $\Delta^{17}$O distributions that we impose, as well as the three different sets of simulation initial conditions that we use. While this is by no means an exhaustive search of parameter space, it suggests that Earth and Venus should be expected to have distinct $\Delta^{17}$O values.

\section{Conclusions}\label{sec:summ}

We have undertaken a statistical study of the giant impact phase of terrestrial planet formation in order to understand the relations between a planet's final mass and orbit and its accretion feeding zone within the initial disk of embryos and planetesimals. We find that the width of a planet's feeding zone does not correlate with its mass or semi-major axis, but is proportional to the width of the initial protoplanetary disk, with massive planets exhibiting remarkably little variance. This means that one should expect terrestrial planets of all sizes to sample the same parts of the disk, provided they form at the same distance from their star.  In other words, super-Earths forming in the habitable zone of their star should have water mass fractions drawn from the same distribution as Earth-size planets.  Moreover, although feeding zones are broad and often overlap in multi-planet systems, the constant feeding zone size dictates that planets forming farther from their star are more likely to accrete water-rich planetesimals from beyond the snow line.

Focusing on planets analogous to our own, we find that Venus analogs are typically built from material orbiting close to the Sun, and Mars analogs are constructed from more distant material, while Earth is a compromise between these two accretion histories. However, these are only general trends. The wings of the feeding zones of each of these planets can span 1--3 AU, depending on the initial planetesimal disk and giant planet orbits we assume. In addition, there is a great deal of stochasticity in the planets' feeding zones, which has a large effect on each planet's inventory of distant, presumably water-rich, planetesimals. The differences in the planets' outer tails of their feeding zones is largely ruled by randomness, and it is not unusual for Venus to have a higher water mass fraction than Mars or Earth.

Our simulations also allow us to study the composition of Theia analogs relative to Venus, Earth, and Mars analogs, and this is a major goal of our work. We find that the feeding zones of Theia analogs vary wildly compared to the planets, and it cannot be assumed that Theia had a similar accretion history to Earth or any of the other planets. Consequently, it is unlikely that Theia's oxygen isotope composition is similar to the Earth's. Although we do not know the distribution of $\Delta^{17}$O among our solar system's initial planetesimals and embryos, we impose four different assumed $\Delta^{17}$O distributions designed to reproduce the observed $\Delta^{17}$O of Mars. Regardless of which family of $\Delta^{17}$O distributions we choose, we find that the probability of a Theia with $|\Delta^{17}$O$| < 0.016$\permil~ is $\sim$5\% or less. Thus, a Theia with an oxygen isotope composition distinctly different from Earth's is the expected outcome of terrestrial planet formation for any of the $\Delta^{17}$O distributions we employ. A consequence of this is that our simulations typically predict that the Moon cannot contain more than $\sim$5--15\% of Theia material, depending on our simulation set and our assumed initial distribution of $\Delta^{17}$O.  There may be more complex $\Delta^{17}$O distributions that yield higher allowable Theia mass fractions for the Moon, but this is not the case for the relatively simple distributions that we employ.

We focus on Theia's oxygen composition since it has been the subject of so much past work, but the Moon is also known to be isotopically similar to the Earth for tungsten, silicon, chromium, and titanium \citep{touboul07, arm12, zhang12}. Presumably, the isotopes of these other elements all had their own unique distributions in the protoplanetary disk that were different from oxygen. While we have shown that Earth and Theia can have feeding zones that by chance yield similar oxygen isotope compositions, the probability that these feeding zones would also simultaneously yield similar isotope compositions for other elements is presumably much smaller. 

In addition, we examine the mass ratios between Theia analogs and the proto-Earth analogs they strike in our simulations. If Theia's mass is at least $\sim$66\% that of the proto-Earth \citet{canup12} demonstrates that the collision of these two bodies can yield a moon with an Earth-like isotope signature even if the two bodies had distinctly different initial compositions. However, we find that such collisions are extremely rare or absent in the final phase of terrestrial planet formation. With over 100 collisions between Theia analogs and proto-Earth analogs, the largest mass ratio we observe is 0.48, and the large majority of collisions have mass ratios below 0.4. 

Another scenario that can produce a terrestrial composition for the Moon is a higher velocity impact between the proto-Earth and Theia \citep{cukstew12,reufer12}. In one set of our simulations we examine the distributions of impact velocities between Earth and Theia analogs, and we find that for Theia analogs with $m>0.1$ M$_{\oplus}$ the typical impact velocity is just a few percent greater than the mutual escape velocity. Meanwhile, most high velocity collisions that successfully reproduce the moon require an impact speed that is at least $\sim$50\% greater than the escape speed. If we consider Theia analog masses down to 0.025 M$_{\oplus}$ there are a significant number of collisions above 150\% of the escape speed, but this is still not the most common outcome. Furthermore, the proto-Earth must presumably be spun up by a massive collision before the moon-forming impact, and we find these massive collisions to be rare. This suggests that a high velocity moon-forming collision may also not be a high probability event. 

Our work demonstrates that although the giant impact hypothesis is the most favored scenario for the origin of the moon, the isotopic similarity between the Earth and the Moon conflicts with predictions of terrestrial planet formation simulations using the simple $\Delta^{17}$O distributions employed here. One potential solution to this is that the outer Earth and moon-forming disk underwent substantial mixing immediately after Theia's impact \citep{pahlstev07}. However, as \citet{canup13} points out, in this scenario one would expect larger isotopic differences between the Earth and the Moon to exist for more refractory elements. Such a trend has not been shown at present. As a result, all explanations for the Moon and Earth's isotopic similarities seem to rely on low probability events, and perhaps our Moon is a statistical outlier in the context of giant impacts.

Finally, using the oxygen isotopic distributions we have generated to study Theia analogs, we also make predictions for the composition of Venus analogs in our simulations. At present there are no known samples of Venus material. However, the oxygen isotope distributions we invoke to explain the isotopic signature of Mars predict that Venus will also have an isotope composition distinct from Earth, regardless of which initial isotope distribution we choose. The $|\Delta^{17}$O$|$ values we predict for Venus are comparable to or greater than those observed for Mars, depending on our assumed initial $\Delta^{17}$O distribution.

\section{Acknowledgements}

This work was partially funded by NSF Grant AST-1109776. We are very grateful to John Chambers, Sean Raymond, and Erik Petersen for helpful discussions while preparing our work for publication. In addition, comments and suggestions from the two anonymous referees greatly improved this paper.

\bibliographystyle{apj}
\bibliography{feeding}

\end{document}